%% file: corelensing.tex
\documentclass[a4paper,11pt]{article}
\pdfoutput=1 

\usepackage{jcappub} 

\usepackage{color}
\input macros.tex

\newcommand{\mission}{\textit{CORE}}
\newcommand{\coremfive}{\mission}
\newcommand{\Euclid}{\textit{Euclid}}

\title{Exploring cosmic origins with \mission: gravitational lensing of the CMB}

\author[1,2,3]{Anthony Challinor,}
\affiliation[1]{Institute of Astronomy, Madingley Road, Cambridge CB3 0HA, UK}
\affiliation[2]{Kavli Institute for Cosmology Cambridge, Madingley Road,
Cambridge CB3 0HA, UK}
\affiliation[3]{DAMTP, Centre for Mathematical Sciences,
Wilberforce Road, Cambridge CB3 0WA, UK}

\author[1,2]{Rupert Allison,}

\author[4]{Julien Carron,}
\affiliation[4]{Department of Physics and Astronomy, University of
Sussex, Falmer, Brighton BN1 9QH, UK}

\author[5]{Josquin Errard,}
\affiliation[5]{Institut Lagrange, LPNHE, Place Jussieu 4, 75005
   Paris, France}

\author[6]{Stephen Feeney,}
\affiliation[6]{Center for Computational Astrophysics, 160 5th
Avenue, New York, NY 10010, USA }

\author[7]{Thomas Kitching,}
\affiliation[7]{ Mullard Space Science Laboratory, University College
London, Holmbury St Mary, Dorking, Surrey RH5 6NT, UK }

\author[8]{Julien Lesgourgues,}
\affiliation[8]{ Institute for Theoretical Particle Physics and
Cosmology (TTK), RWTH Aachen University, D-52056 Aachen, Germany. }

\author[4]{Antony Lewis,}

\author[1,2]{\'{I}\~{n}igo Zubeld\'{i}a,}

\author[9,10]{Ana Achucarro,}
\affiliation[9]{Instituut-Lorentz for Theoretical Physics,
Universiteit Leiden, 2333 CA, Leiden, The Netherlands}
\affiliation[10]{Department of Theoretical Physics, University of the
Basque Country UPV/EHU, 48040 Bilbao, Spain}

% ...

\emailAdd{a.d.challinor@ast.cam.ac.uk}

\abstract{Lensing of the cosmic microwave background (CMB) is now a
  well-developed probe of the clustering of the large-scale mass
  distribution over a broad
  range of redshifts. By exploiting the non-Gaussian imprints of
  lensing in the polarization of the CMB, the \mission\ mission will
  allow production of a clean map of the lensing deflections over
  nearly the full-sky. The number of high-$S/N$ modes in this map will
  exceed current CMB lensing maps by a factor of 40, and the
  measurement will be sample-variance limited on all scales where
  linear theory is valid. Here, we summarise this mission product and
  discuss the science that will follow from its power spectrum
  and the cross-correlation with other clustering data. For example,
  the summed mass of neutrinos will be determined to an accuracy of
  $17\,\text{meV}$ combining \mission\ lensing and CMB two-point
  information with contemporaneous measurements of the baryon acoustic oscillation
  feature in the clustering of galaxies, three times smaller than
  the minimum total mass allowed by neutrino oscillation
  measurements. Lensing has applications across many other science
  goals of \mission, including the search for $B$-mode polarization
  from primordial gravitational waves. Here, lens-induced $B$-modes
  will dominate over instrument noise, limiting constraints on the
  power spectrum amplitude of primordial gravitational waves. With
  lensing reconstructed by \mission, one can ``delens'' the observed
  polarization internally, reducing the lensing $B$-mode power by
  60\,\%. This can be improved to 70\,\% by combining lensing and
  measurements of the cosmic infrared background from \mission,
  leading to an improvement of a factor of $2.5$ in the error on the
  amplitude of primordial gravitational waves compared to no delensing
  (in the null hypothesis of no primordial $B$-modes). Lensing
  measurements from \mission\ will allow calibration of the halo masses of 
  the tens of thousands of galaxy clusters that it will find, with
  constraints dominated by the clean polarization-based estimators. The 19
  frequency channels proposed for \mission\ will allow accurate
  removal of Galactic emission from CMB maps. We present initial
  findings that show that residual Galactic foreground contamination
  will not be a significant source of bias for lensing power spectrum
  measurements with \mission.}

\begin{document}
\maketitle
\flushbottom

\section{Introduction}
\label{sec:intro}

The cosmic microwave background (CMB) is gravitationally lensed by
large-scale structure as it propagates from the last-scattering
surface, leading to a subtle remapping of the temperature and polarization 
anisotropies (e.g., Ref.~\cite{Lewis:2006fu}). Lensing imprints information in the CMB about
the geometry of our Universe and the late-time clustering of matter. This information is 
otherwise degenerate in the primary CMB fluctuations that are
generated at last scattering~\cite{Hu:2001fb}. The lensing deflection field can be reconstructed 
using sensitive, high-resolution observations, potentially providing a large-scale, nearly full-sky map of the 
integrated mass in the entire visible Universe. The power spectrum of
this map, when combined with the power spectra of the temperature and
polarization anisotropies, constrains parameters such as the (summed) mass
of neutrinos and spatial curvature using the CMB
alone~\cite{2011PhRvL.107b1302S,Ade:2015zua}. Lensing is sensitive to all matter along the line of sight, and not
just the luminous matter probed, for example, by galaxy redshift
surveys. CMB lensing is therefore highly complementary to other
tracers of large-scale structure. For instance, by cross-correlating
one can calibrate the astrophysical and instrumental bias relations between the 
tracers and the underlying density field, which is critical to maximize the 
returns from future surveys (see e.g.,
Ref.~\cite{2012ApJ...759...32V}). Furthermore, the reconstructed
lensing map can be used to remove partly the effects of lensing, which
would otherwise obscure our view of the primary fluctuations. A particularly
important application of such ``delensing'' is in the search for
primordial gravitational waves via large-angle $B$-mode
polarization, where it can provide critical improvements in primordial
constraints~\cite{Kesden:2002ku,Knox:2002pe,Seljak:2003pn}.

In the past decade, CMB lensing has gone from its first
detection~\cite{Smith:2007rg,Das:2011ak} to becoming a well-established, precision probe
of clustering. Reconstructed maps of the CMB lensing deflections have
been made with data from ground-based instruments (e.g., Refs.~\cite{Ade:2013gez,Story:2014hni,Sherwin:2016tyf,B2Keck:2016afx,Omori:2017tae})
and from the \planck\
satellite~\cite{Ade:2013zuv,Ade:2015zua}. Due to its nearly full-sky coverage, the \planck\
lensing results currently have the greatest statistical
power but are very far from exhausting the information available in
the lensed CMB. For this reason, lensing is a major goal being
targeted by nearly all forthcoming and proposed experiments. These
include the Cosmic Origins Explorer (\mission), a satellite mission
recently proposed to the European Space Agency's fifth call for a medium-class mission.

This paper is one of a series written as part of the development of
the \mission\ mission concept and science case. Here, we describe how
a full-sky CMB lensing map can be reconstructed with \mission\ data,
quantify the expected statistical precision of this map, and
illustrate its application across several of the key science targets
of the mission. Most of our forecasted results are presented for the
baseline mission concept, described in detail elsewhere in this
series~\cite{Delabrouille:2017}. However, in some places we present parametric
comparisons of different options to justify design choices made in the
baseline. Lensing impacts much of \mission\ science; closely related papers in
this series describe constraints on inflation~\cite{Finelli:2016cyd} (where
delensing is significant), cosmological
parameters~\cite{DiValentino:2016foa} (which combines the temperature, polarization,
and lensing power spectra), and galaxy cluster science~\cite{Melin:2017lkr}
(where mass calibration with CMB lensing of temperature anisotropies
is discussed).

This paper is organised as follows. Section~\ref{sec:recon} introduces
lensing reconstruction and reviews the current observational
status. Some further technical details are summarised in Appendix~\ref{app:quadrecon}.
Lensing reconstruction with \mission\ is described in
Sec.~\ref{sec:corelensing}. CMB lensing is expected to be a particularly clean
probe of the absolute mass scale of neutrinos, through the impact of
their mass on the growth of cosmic structure; this important target
for \mission\ is discussed in Sec.~\ref{sec:mnu}. In Sec.~\ref{sec:LSS}, we
outline the complementarity between CMB lensing and other tracers of
large-scale structure and forecast the improvements that would arise
from combining lensing from \mission\ data with contemporaneous large-scale
structure surveys. Delensing of $B$-mode polarization is discussed in
Sec.~\ref{sec:delens} and the implications for constraining primordial
gravitational waves are reviewed. In Sec.~\ref{sec:clustermass} we highlight the
potential for \mission\ to self-calibrate the
masses of its cluster catalogue via lensing of CMB polarization,
extending the temperature-based forecasts presented in
Ref.~\cite{Melin:2017lkr}. While most of the forecasts throughout this paper assume
that the 19 frequency channels of \mission\ will allow accurate cleaning of
Galactic foreground emission, and so ignore potential Galactic
residuals, in Sec.~\ref{sec:galforegrounds} we relax this assumption. We
present initial results, based on
simulated maps of the polarized Galactic dust emission, on the bias
that can arise in the lensing power spectrum from temperature- and
polarization-based reconstructions in the pessimistic scenario that
dust cleaning is ineffective. 

\section{CMB lensing reconstruction}
\label{sec:recon}

\begin{figure}
\begin{center}
\includegraphics[width=7cm,angle=-90]{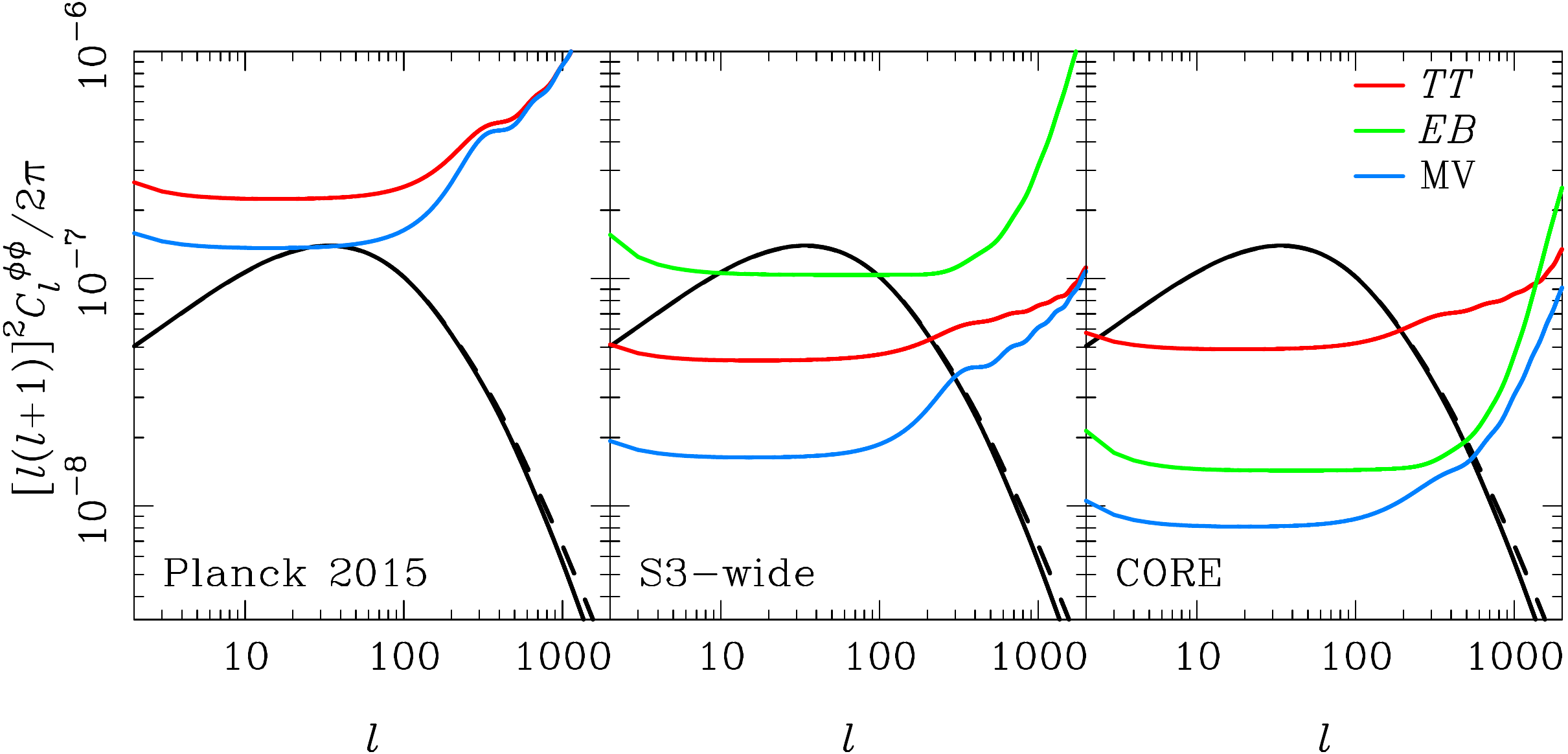}
\end{center}
\caption{Reconstruction noise of the lensing deflection power spectrum
from \textit{Planck} 2015 (left) and as forecast for S3-wide (middle)
and \mission\ (right). S3-wide represents a third-generation wide-area
(sky fraction of around 40\,\%)
ground-based experiment, with specifications similar to AdvACT. In
particular, we follow~\cite{Allison:2015qca} by assuming a beam size of
$1.4\,\text{arcmin}$, a temperature sensitivity of
$8.0\,\mu\text{K}\,\text{arcmin}$ and polarization sensitivity of $11.3\,\mu\text{K}\,\text{arcmin}$.
The deflection power spectrum is plotted based on the linear matter power spectrum (black
solid) and with nonlinear corrections (black dashed).
}
\label{fig:Nlzero}
\end{figure}

Lensing by large-scale structure remaps the CMB temperature and
polarization fluctuations imprinted on the last-scattering
surface. The lenses lie at all redshifts back to last-scattering, but
the peak lensing efficiency is around $z=2$. Large-scale lenses, with
$k \lesssim 0.01\,\text{Mpc}^{-1}$, dominate the lensing signal except
on the smallest angular scales, making CMB lensing a particularly powerful
probe of $O(100) \,\text{Mpc}$ structures at high redshift. The
lensing deflections are small, with r.m.s.\ of
$2.5\,\text{arcmin}$, but are coherent over several degrees. To an
excellent approximation, the deflection field can be expressed as the
angular gradient of the \emph{CMB lensing potential} $\phi$, which is
itself an integral of the $3D$ gravitational potential
along the (background) line of sight. The angular power spectrum of
the deflection field, $l(l+1)C_l^{\phi\phi}$, is shown in Fig.~\ref{fig:Nlzero}.

Lensing
has several observable effects on the CMB (see Refs.~\cite{Lewis:2006fu,2010GReGr..42.2197H} for reviews).
It smooths out the acoustic peaks in
the temperature and $E$-mode polarization power spectra and transfers
power from large to small scales. This peak smoothing is routinely
included when deriving cosmological parameter constraints from the CMB
power spectra and the effect itself is detected at more than
$10\,\sigma$ in the measurements of the $TT$ power spectrum from \Planck~\cite{Ade:2015xua}.
Lensing also partially converts $E$-mode polarization into
$B$-mode~\cite{Zaldarriaga:1998ar}. These lens-induced $B$-modes have an almost
white-noise spectrum, corresponding to around
$5\,\mu\text{K}\,\text{arcmin}$ of noise, on the large angular scales relevant
for searches for primordial $B$-mode polarization sourced by a
stochastic background of primordial gravitational waves (see
Sec.~\ref{sec:delens}). Finally, lensing induces non-Gaussianity in the
CMB, which shows up as higher-order non-zero connected moments (in
particular the trispectrum or connected 4-point function) and as
non-zero 3-point correlator between pairs of CMB fields and tracers of
large-scale structure~\cite{Zaldarriaga:2000ud,Hu:2001fa}.

Exploitation of the non-Gaussianity induced by lensing can be
conveniently thought of as a two-step process. The first involves
\emph{lens reconstruction}, whereby an estimate for the lensing
potential $\phi$ is obtained from quadratic combinations of the
observed CMB fields~\cite{Hu:2001tn}. In the second step, the lens
reconstruction is correlated with itself, to estimate the lensing
potential power spectrum $C_l^{\phi\phi}$, or an external tracer of
large-scale structure, to estimate the correlation between the lensing
potential and the tracer. The process of lens reconstruction can be understood by noting
that for fixed $\phi$, lensing induces anisotropic 2-point correlations in the
CMB. The linear response of the covariance between lensed CMB fields
$\tilde{X}_{l m}$ and $\tilde{Y}_{lm}$, where $X$ and $Y = T$, $E$, or $B$, to a variation
in the lensing potential is
\begin{equation}
\langle \delta (\tilde{X}_{l_1 m_1} \tilde{Y}_{l_2 m_2})
\rangle \approx \sum_{LM} (-1)^M \left(
\begin{array}{ccc}
l_1 & l_2 & L \\
m_1 & m_2 & -M
\end{array}
\right) \mathcal{W}^{XY}_{l_1 l_2 L} \delta \phi_{LM} \, ,
\end{equation}
where the covariance response functions $\mathcal{W}^{XY}_{l_1 l_2 L} $ are given
in Appendix~\ref{app:quadrecon} (see also Ref.~\cite{Okamoto:2003zw}). An optimal quadratic estimator $\hat{\phi}_{LM}$ can be
written in the form\footnote{Generally, it is necessary also to subtract a
  \emph{mean field} term from the estimator to deal with survey
  anisotropies such as masking and anisotropic instrument noise and
  filtering.}
\begin{equation}
\hat{\phi}^{XY}_{LM} = \frac{(-1)^M}{2}\frac{1}{\mathcal{R}_L^{XY}} \sum_{l_1 m_1,
  l_2 m_2} \left(
\begin{array}{ccc}
l_1 & l_2 & L \\
m_1 & m_2 & -M
\end{array}
\right)[\mathcal{W}^{XY}_{l_1 l_2 L}]^\ast \bar{X}_{l_1 m_1} \bar{Y}_{l_2
  m_2} \, ,
\label{eq:quadest}
\end{equation}
where $\bar{X}$ and $\bar{Y}$ are the inverse-variance filtered fields
and the normalisation $\mathcal{R}_L^{XY}$ is chosen to ensure the
estimator is unbiased. The individual quadratic estimators can be
combined linearly to give a minimum-variance (MV) combination:
$\hat{\phi}_{LM}^{MV} = \sum_{XY} \hat{\phi}_{LM}^{XY}
\mathcal{R}_L^{XY}/ \sum_{XY} \mathcal{R}_L^{XY}$.

Lens reconstruction is statistical, with Gaussian fluctuations of the
CMB giving rise to a statistical noise in the reconstruction. This
reconstruction noise is similar to shape noise in galaxy lensing,
whereby the intrinsic ellipticity of a galaxy adds white noise to the
estimated gravitational shear. The reconstruction noise can be
quantified by its power spectrum, usually denoted
$N_L^{(0)}$. Consider forming the power spectrum of
$\hat{\phi}_{LM}^{XY}$. This is quartic in the CMB fields and the
connected part of this 4-point function gives simply $C_L^{\phi\phi}$
(plus an additional non-local coupling to the potential power spectrum,
$N_L^{(1)}$, which arises from non-primary
couplings~\cite{Hu:2001fa,Cooray:2002py}) while the disconnected part gives
$N_L^{(0)}$. The lens reconstruction has high $S/N$ on scales where
$C_L^{\phi\phi} \gg N_L^{(0)}$. Examples of $N_L^{(0)}$ for various experiments are given
in Fig.~\ref{fig:Nlzero}.

\begin{figure}
\begin{center}
\includegraphics[width=12cm,angle=0]{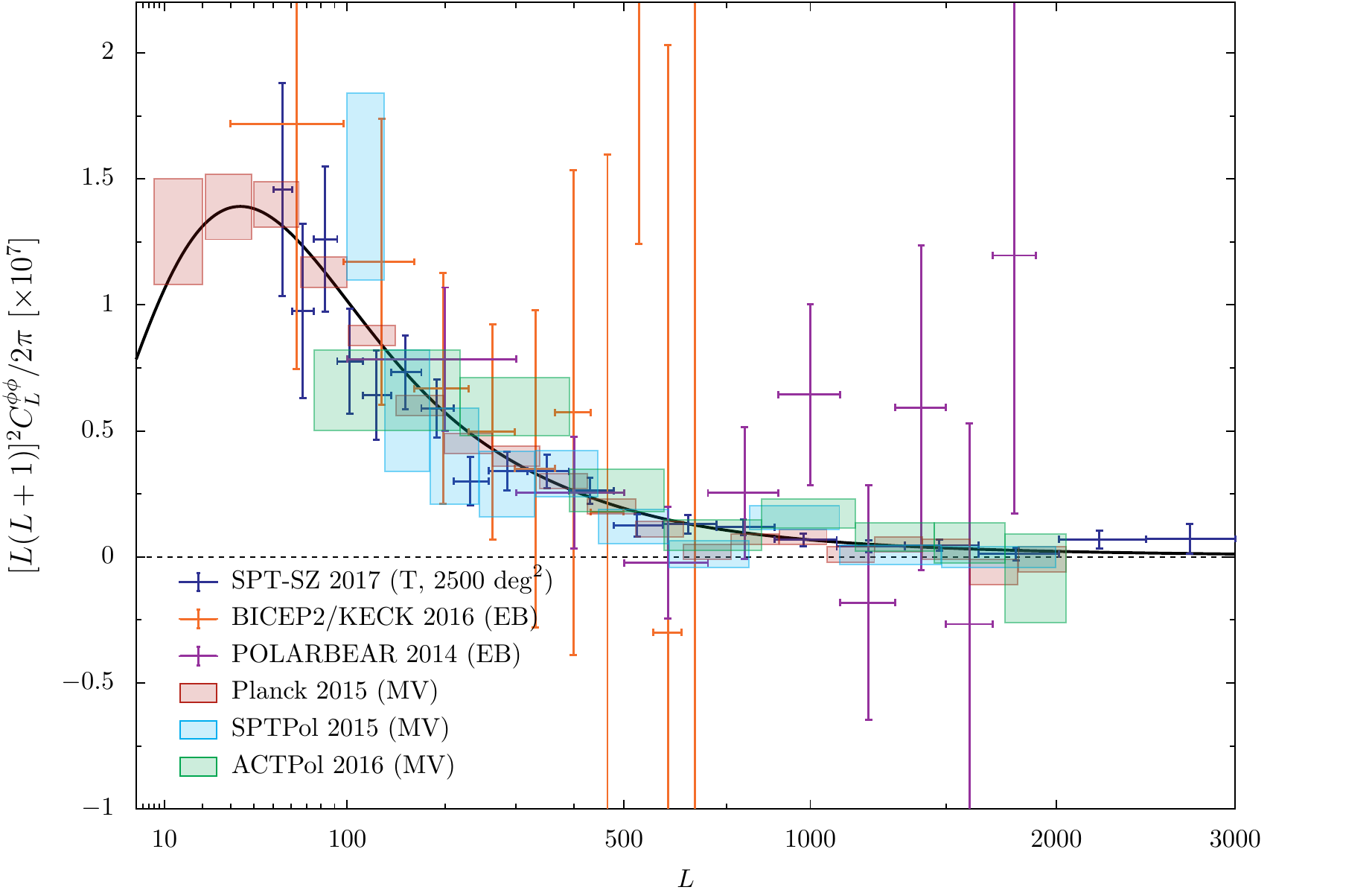}
\end{center}
\caption{Current lensing potential power spectrum measurements from
  \textit{Planck} 2015~\cite{Ade:2015zua},
  SPTpol~\cite{Story:2014hni}, POLARBEAR~\cite{Ade:2013gez},
  ACTPol~\cite{Sherwin:2016tyf}, BICEP2/Keck
  Array~\cite{B2Keck:2016afx}, and SPT-SZ~\cite{Omori:2017tae}.}
\label{fig:lensingresults}
\end{figure}

CMB lensing is a rapidly advancing frontier of observational
cosmology. Estimates of the lensing potential power spectrum from the CMB
4-point function from \Planck, and several ground-based experiments,
are shown in Fig.~\ref{fig:lensingresults}. The \Planck\ results~\cite{Ade:2015zua} provide the highest $S/N$ 
detection of CMB lensing to date (around $40\,\sigma$). At the noise
levels of \Planck\ (around $30\,\mu {\rm 
K}\,{\rm arcmin}$ in temperature), the $TT$ estimator has the highest $S/N$ and
dominates the MV combination, as shown in the left-hand panel of Fig.~\ref{fig:Nlzero}.
On large angular scales, the reconstruction noise power is
approximately~\cite{2011PhRvD..83d3005H}
\begin{equation}
[L(L+1)]^2 N_L^{(0)} \approx \left\{ \frac{1}{8} \sum_l \frac{2l+1}{4\pi}
  \left(\frac{C_l^{TT}}{C_{l,\text{tot}}^{TT}}\right)^2\left[\left(\frac{d\ln
        \mathcal{D}^{TT}_l}{d\ln l}\right)^2 +
    \frac{1}{2}\left(\frac{d\ln C_l^{TT}}{d\ln l}\right)^2\right]\right\}^{-1}\, ,
\label{eq:TTlargelens}
\end{equation}
where $C_l^{XY}$ is the (lensed) CMB power spectrum between fields $X$
and $Y$, $C_l^{XY}$ is the total spectrum including (beam-deconvolved)
instrument noise for $X=Y$, and $\mathcal{D}_l^{XY} \equiv l(l+1)C_l^{XY}/(2\pi)$.
The power spectrum $L^2(L+1)^2 N_L^{(0)}/4$ is
approximately constant on large scales corresponding to white
noise in the reconstructed convergence ($\kappa = - \nabla^2 \phi
/2$) or shear ($\gamma = - \eth^2 \phi /2$).
This behaviour arises since for large-scale lenses, the convergence
and shear are reconstructed locally from much smaller-scale CMB
anisotropies. The convergence produces dilation of the local
small-scale CMB power spectrum, while the shear produces local anisotropy.
It can be shown that the term involving $d\ln
\mathcal{D}_l^{TT}/d\ln l$ in Eq.~(\ref{eq:TTlargelens}) is the
information from the convergence (and so vanishes for a
scale-invariant spectrum $\mathcal{D}^{TT}_l = \text{const.}$), while
the term involving $d\ln C_l^{TT}/d\ln l$ is the information from the shear~\cite{2012PhRvD..85d3016B}.

While the $S/N$ of the $TT$ estimator can be improved by increasing
the resolution and sensitivity beyond \Planck, it can never exceed
unity for scales smaller than multipole $L \approx 200$. Furthermore,
extragalactic foregrounds make using the temperature anisotropies
very difficult at scales $l > 2500$. Rather, the way to improve lensing reconstructions
significantly is to use 
high-sensitivity polarization observations~\cite{Hu:2001kj}. In
particular, if the lens-induced $B$ modes can be mapped with high
$S/N$, the $EB$ estimator becomes the most powerful. On large angular
scales, the reconstruction noise power for this estimator is
approximately (for $L\geq 2$)
\begin{equation}
L^4 N_L^{(0)} \approx \left( \frac{1}{2} \sum_l \frac{2l+1}{4\pi}
  \frac{(C_l^{EE})^2}{C_{l,\text{tot}}^{EE} C_{l,\text{tot}}^{BB}}
\right)^{-1} \, ,
\label{eq:EBlargelens}
\end{equation}
and is limited by the total $B$-mode power $C_{l,\text{tot}}^{BB}$
(including instrument noise),
which can be very small for low-noise observations. The $EB$ estimator
for large scale lenses is only sensitive to shear as the dilation of
small-scale polarization by a constant convergence does not convert
$E$-mode polarization into $B$-mode.\footnote{Indeed, the reconstruction noise
on the $EB$ estimator is very large at $L=1$ since the dipole of the
lensing potential produces no shear.}
Polarization-based lens reconstructions have been demonstrated 
recently from ground-based
experiments~\cite{Ade:2013gez,Story:2014hni,B2Keck:2016afx,Sherwin:2016tyf}, and also \emph{Planck}, but are currently very noisy. Future, funded
wide-area CMB surveys (see S3-wide in Fig.~\ref{fig:Nlzero},
which has specifications similar to AdvACT) also do not have the
sensitivity to exploit polarization-based lensing fully. To image the
lens-induced $B$-modes requires the polarization noise level to be well below
$5\,\mu\text{K}\,\text{arcmin}$. Achieving such sensitivity over a
large fraction of the sky -- to maximise the number of resolved lensing
modes and the overlap with large-scale structure surveys -- would
require a ground-based experiment with around $5\times 10^5$
detectors. Plans for such a programme, CMB-S4, are currently under
development~\cite{Abazajian:2016yjj}. Alternatively, the same goal can be reached
with almost two orders of magnitude fewer detectors from a space-based
experiment, as we discuss in Sec.~\ref{sec:corelensing}.

Finally, we note that at very low noise levels it is possible to
improve over lens reconstructions based on quadratic estimators
(e.g., Refs.~\cite{Hirata:2002jy,Hirata:2003ka,Anderes:2014foa,Carron:2017mqf}). For
example, we see from Eq.~(\ref{eq:EBlargelens}) that the precision of
the $EB$ estimator is limited at low noise levels by the small-scale
lens-induced $B$-mode power. However, simple field counting suggests
that with no noise we should be able to invert the observed $E$- and
$B$-fields to recover the unlensed $E$-modes and the lensing potential
$\phi$. For the noise levels of \mission, the improvement from more
optimal estimators is rather modest and so, for simplicity, most of
the forecasts in this paper are based on quadratic
estimators. However, in Sec.~\ref{sec:delens} we do discuss further
the improvements in constraints on primordial gravitational waves that
arise from delensing with a more optimal lens reconstruction.

\section{Lens reconstruction with \mission}
\label{sec:corelensing}

The baseline configuration for the \mission\ mission is summarised in
Table~1 of Ref.~\cite{Delabrouille:2017}. Briefly, it consists of 19 frequency
channels in the range 60--600\,GHz with beam sizes (full width at
half-maximum) ranging from 18\,arcmin (at 60\,GHz) to 2\,arcmin (at
600\,GHz). For the forecasts in this paper, we combine the six
channels in the frequency range 130--220\,GHz with
inverse-variance noise weighting, assuming that the channels outside
this range can be used to clean Galactic foregrounds without further
significant loss of sensitivity. The polarization sensitivity of each
of the six ``CMB'' channels is around $5\,\mu {\rm 
K}\,{\rm arcmin}$ in polarization (and a factor $\sqrt{2}$ better in
temperature) assuming a four-year mission. The combination of the CMB
channels gives a polarization sensitivity of $2.1\,\mu {\rm 
K}\,{\rm arcmin}$ and an effective resolution of around
$6.2$\,arcmin. 

\begin{figure}
\begin{center}
\includegraphics[width=8cm,angle=-90]{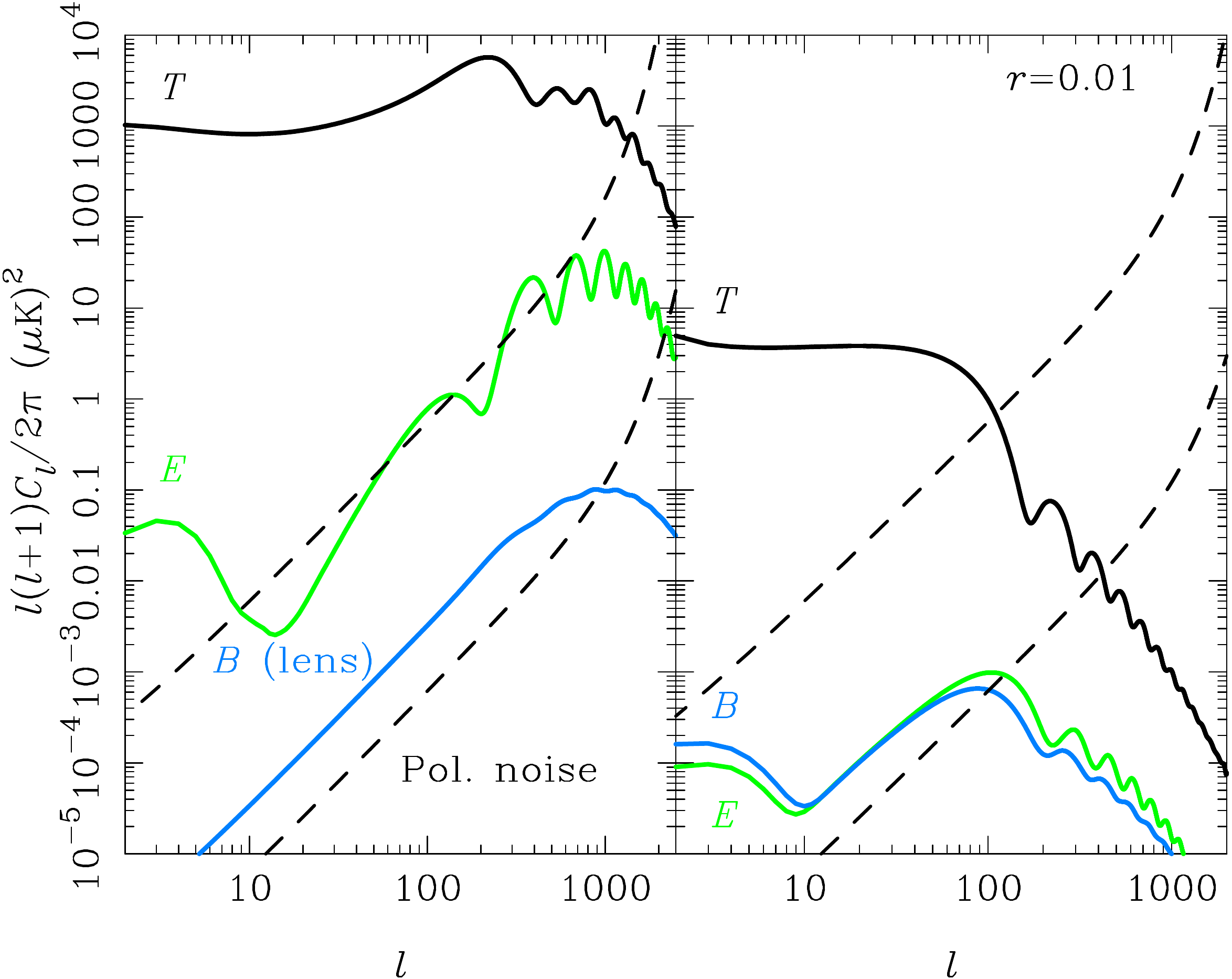}
\end{center}
\caption{Power spectra of the polarization noise for \mission\
  (lower dashed lines) and \textit{Planck} 2015 (upper dashed lines) compared to
  the $TT$ (black), $EE$ (green), and $BB$ (blue) power spectra from
  curvature perturbations (left) and gravitational waves for $r=0.01$ (right).
}
\label{fig:noisepower}
\end{figure}

The polarization noise power spectrum of the combination of the six
CMB channels is shown in Fig.~\ref{fig:noisepower}, where it is
compared to the CMB $TT$, $EE$, and $BB$ power spectra from curvature
fluctuations and from primordial gravitational waves with a
tensor-to-scalar ratio $r=0.01$. We see that with \mission, the
$E$-mode polarization has $S/N>1$ for multipoles $l<2000$ and the
lens-induced $B$-modes have $S/N>1$ for
$l<1000$. Figure~\ref{fig:noisepower} also compares the noise power to
that of the full \textit{Planck} survey; \mission\ has around 30 times
the polarization sensitivity of \textit{Planck}.

The noise levels $N_L^{(0)}$ on lens reconstructions from \mission\ in its
baseline configuration are shown in the right-hand panel of
Fig.~\ref{fig:Nlzero} for a temperature-based quadratic estimator, the
$EB$ estimator, and the minimum-variance combination of all five
quadratic estimators. The $EB$ estimator is the most powerful
quadratic estimator since, as noted above, \mission's polarization
sensitivity of $2.1\,\mu {\rm K}\,{\rm arcmin}$ and angular resolution
allow imaging of the lens-induced $B$-modes. This situation is quite
different from \textit{Planck}, and from the current generation of
wide-area surveys (see S3-wide in Fig.~\ref{fig:Nlzero}). For these,
lensing reconstruction is dominated by the $TT$ estimator. This
transition to the regime where $EB$ dominates is transformational for
two reasons. First, only then is it possible to achieve high $S/N$
reconstructions of lenses at multipoles $L>200$ and so maximise the
cosmological information that can be extracted from CMB
lensing. Second, the non-Gaussian nature of extragalactic foregrounds
to the temperature anisotropies (e.g., radio and infrared galaxies and the thermal Sunyaev-Zel'dovich
signal from galaxy clusters) can bias estimation of the lensing power
spectrum and generally requires correction~\cite{vanEngelen:2013rla}.
However, lens reconstructions based on
polarization are expected to be much cleaner than those from
temperature~\cite{Smith:2008an}.

\begin{figure}
\begin{center}
\includegraphics[width=8cm,angle=-90]{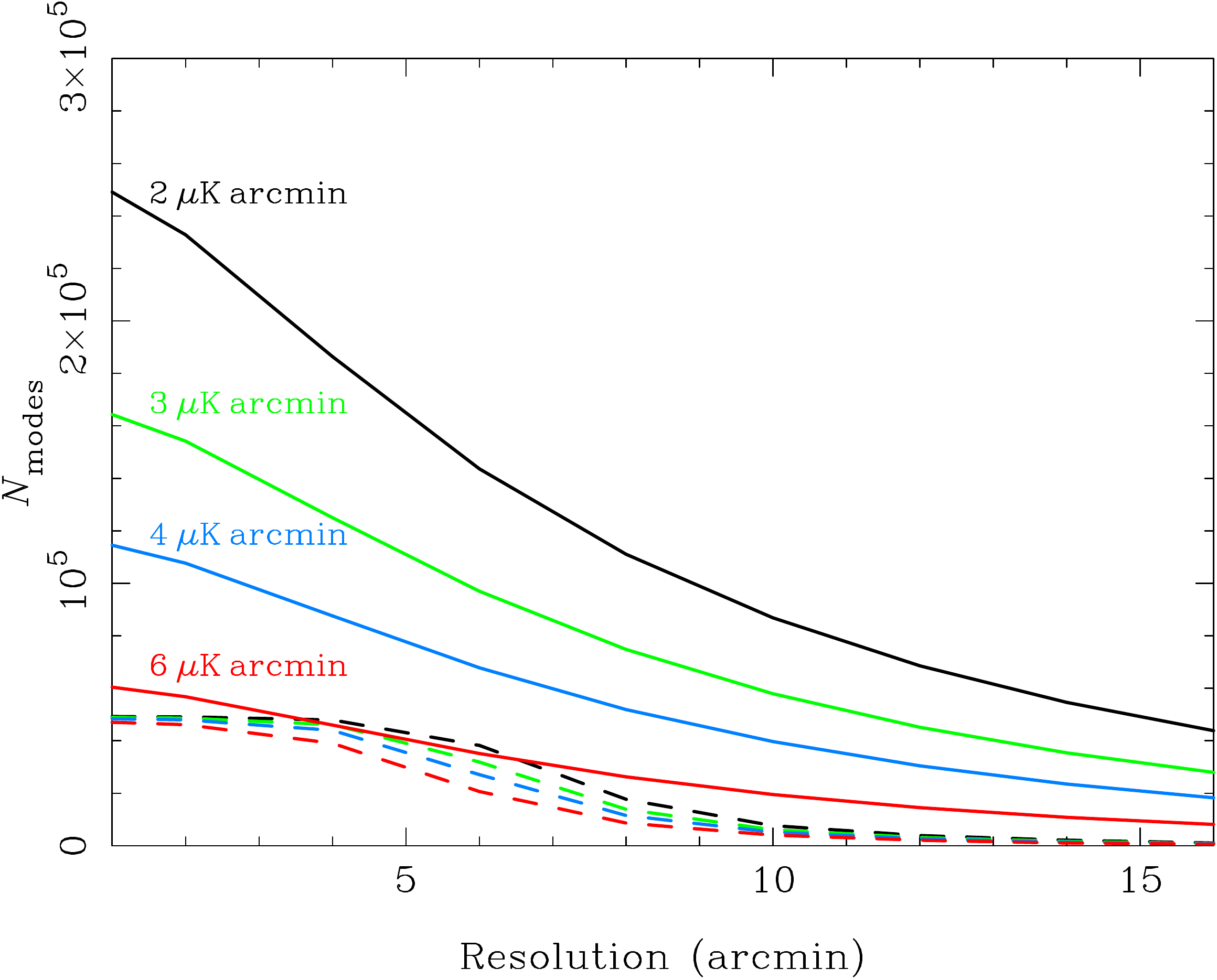}
\end{center}
\caption{Number of effective resolved lensing reconstruction modes as a function of
  angular resolution for surveys covering 70\,\% of the sky for the
  indicated polarization noise levels. The solid lines are for the
  $EB$ quadratic estimator while the dashed lines are for $TT$. In all
  cases, CMB modes are only used up to $l_{\text{max}} = 3000$ in the
  quadratic estimators.}
\label{fig:Nmodes}
\end{figure}

We see from Fig.~\ref{fig:Nlzero} that \mission\ will reconstruct lensing with $S/N>1$ 
\emph{per mode} up to multipoles $L \approx 550$ 
over nearly the full sky. Significantly, \coremfive\ can extract
essentially all of the information in the lensing power
spectrum on scales where linear theory is reliable. A useful way to summarise the information
content of the lens reconstruction is through the total $S/N$ of a
measurement of the amplitude of the lensing power spectrum, i.e.,
\begin{equation}
\left(\frac{S}{N}\right)^2 \approx f_{\text{sky}} \sum_L \frac{2L+1}{2}
\left(\frac{C_L^{\phi\phi}}{C_L^{\phi\phi} + N_L^{(0)}}\right)^2 \, ,
\end{equation}
where $f_{\text{sky}}$ is the fraction of the sky that is usable for
lensing science with the survey. Based on experience with
\textit{Planck}, we expect $f_{\text{sky}} \approx 0.7$ for
\mission. Note that $(S/N)^2$ is just half the effective number of
modes in the reconstruction and so we define $N_\text{modes} \equiv 2
(S/N)^2$. For \mission, $N_\text{modes} \approx 1.6\times 10^5$; for comparison, 
\begin{equation}
N_{\text{modes}} = \begin{cases}
4.0\times 10^3 & \text{\Planck\ 2015} \\
3.9\times 10^4 & \text{S3-wide} \\
1.6\times 10^5 & \text{\mission} \, ,
\end{cases}
\end{equation}
assuming S3-wide can use 40\,\% of the sky. Figure~\ref{fig:Nmodes}
shows $N_{\text{modes}}$ for 70\,\% sky coverage as a function of
angular resolution for polarization noise levels in the range 2--6\,$\mu {\rm 
K}\,{\rm arcmin}$. For polarization noise better than $5\,\mu{\rm
K}\,{\rm arcmin}$ (i.e., levels where imaging of the lens-induced
$B$-modes becomes possible), the $EB$ estimator indeed dominates
$N_{\text{modes}}$. The number of lensing modes from the $EB$
estimator continues to increase
with decreasing noise levels as $B$-modes on smaller
scales (where the lens-induced $B$-mode power is not white) are
imaged. Note, however, that increasing $N_{\text{modes}}$ does not
necessarily lead to improved parameter constraints from the lensing
spectrum as these can rather be limited by parameter degeneracies (see
Sec.~\ref{sec:mnu} for the case of neutrino masses). 

\section{Absolute neutrino mass scale}
\label{sec:mnu}

Several aspects of the neutrino sector are still not well
understood.
In particular, neutrino oscillations show that neutrinos
must be massive, with the flavour eigenstates a mixture of mass
eigenstates. Oscillations are sensitive to the differences of the
squared masses, but not to the absolute mass scale. Since neutrinos
are so numerous, even small masses can have a significant cosmological
effect making \mission\ a powerful probe of their unknown absolute mass
scale. In addition, the usual assumption that the three active flavour states
(i.e., those that participate
in the weak interaction) mix with
three mass eigenstates has been questioned in light of a number of
anomalies found with short-baseline oscillation and reactor measurements
(see Ref.~\cite{Abazajian:2012ys} for a review). Instead, one or
more additional sterile neutrinos can be introduced, which do not
participate in weak interactions, and that, alongside the active
states, mix with four or more mass eigenstates. Sterile--active mass
splittings at the eV scale are required to resolve the above
anomalies, but are disfavoured by current cosmological bounds (e.g.,
Ref.~\cite{Ade:2015xua}). CMB data are sensitive to the mass of sterile
neutrinos through lensing, while the damping tails of the
temperature and polarization power spectra provide sensitivity to their effective number.

\subsection{Masses of active neutrinos}

Neutrino oscillation data show that neutrinos must be massive, but the
data are insensitive to the absolute neutrino mass
scale. Cosmological observations are naturally complementary since
they are sensitive mostly to the total mass with only weak sensitivity
to the mass splittings. The
mass splittings inferred from oscillations, $m_2^2-m_1^2 = (7.53\pm 0.18)\times 10^{-5}\,{\rm
  eV}^2$ and $|m_3^2-m_2^2| = (2.44\pm 0.06)\times 10^{-3} \,{\rm
  eV}^2$~\cite{Olive:2016xmw}, imply two possible mass orderings: the normal ordering ($m_3
> m_2 > m_1$) with a minimum total mass of $\sum m_\nu \approx
59\,{\rm meV}$; and the inverted ordering ($m_2 > m_1 > m_3$) with a
minimum total mass of $98\,{\rm meV}$. The mass scale can also be
probed kinematically with laboratory $\beta$-decay experiments. At the
target minimal-mass scales, the effective masses that are probed with such
experiments are well below the detection limits of current and future
planned experiments. However, next-generation searches for neutrinoless double
beta decay (which would require neutrinos to be Majorana particles)
are expected to reach sensitivities to the relevant effective mass
that could allow detection if the ordering is inverted (e.g., Ref.~\cite{DellOro:2014ysa}).

Neutrinos with masses less than around $0.5\,\text{eV}$ were still
relativistic around the time of recombination. Their effect on the
\emph{primary} CMB anisotropies is therefore limited to projection effects
due to the change in the angular diameter distance to last
scattering. If we keep the physical densities of CDM, baryons and dark
energy fixed, an increase in the neutrino mass increases the expansion
rate after neutrinos become non-relativistic. The associated reduction in the
angular diameter distance to last scattering can be offset by a
reduction in the dark energy density (or, equivalently, the Hubble constant).
This geometric degeneracy limits our ability to probe lighter neutrino
masses with the primary CMB anisotropies alone; for example, the
95\,\% upper limit on the summed neutrino mass from \Planck\
temperature and polarization anisotropies is $\sum m_\nu <
0.49\,\text{eV}$~\cite{Ade:2015xua}. However, the modification to the
expansion rate affects geometric probes, such as the measurement of
the baryon acoustic oscillation (BAO) feature in the clustering of
galaxies, which can be used to break the CMB geometric degeneracy. For example,
combining \Planck\ with current BAO data improves the constraint to around $\sum
m_\nu < 0.2\,\text{eV}$~\cite{Ade:2015xua,Alam:2016hwk}. In models with
curvature or dynamical dark energy the geometric degeneracy is
further exacerbated and the constraints on $\sum m_\nu$ are weakened.

Massive neutrinos also affect the growth of structure on scales
smaller than the horizon size when neutrinos become non-relativistic,
leaving a
distinctive feature in the lensing potential power spectrum.
Massive neutrinos can only cluster on scales larger than their
free-streaming scale, roughly the product of their r.m.s.\ speed and
the Hubble time. Once neutrinos become non-relativistic, their comoving
free-streaming scale \emph{decreases} with time as their r.m.s.\ speed
falls as $1/a$, where $a$ is the scale factor. For reference, at redshift $z=2$ where the kernel for
CMB lensing peaks, the associated comoving wavenumber is
$k_{\text{fs}} \approx 0.09 (m_\nu/50\,\text{meV})\,\text{Mpc}^{-1}$
-- see, for example, Ref.~\cite{Lesgourgues:2006nd} -- 
corresponding to a
multipole $l \approx 60$ for $m_\nu = 50\,\text{meV}$ (see, e.g., Ref.~\cite{Lesgourgues:2006nd}). 
The increase in the expansion rate due to
non-relativistic massive neutrinos slows the growth of
structure in the other matter components on scales smaller than the
free-streaming scale. At any given redshift, the net effect in the
power spectrum of the gravitational potential is an almost constant
fractional suppression for $k > k_{\text{fs}}(a)$. Scales larger than
the horizon size at the non-relativistic transition are not suppressed
since neutrinos have always clustered on such scales, mitigating the effect of
the enhanced expansion rate on the growth of structure. For a given
mass, the amount of suppression in the lensing
potential power spectrum $C_l^{\phi\phi}$ depends on exactly which
other parameters are held fixed. For example, moving along the
geometric degeneracy of the primary CMB anisotropies (i.e., fixing the
physical densities in CDM and baryons, the angular-diameter distance
to last scattering, the primordial power spectrum and the optical
depth to reionization), $C_l^{\phi\phi}$
is suppressed by around 1.5\,\% for a total mass $\sum m_\nu =
0.06\,\text{eV}$ compared to the massless case. By way of comparison,
the amplitude of the lensing potential power spectrum can be measured
with a $1\,\sigma$ error of around $0.35\,\%$ with \mission, although,
as we shall see below, this does not translate directly into a
constraint on the summed neutrino mass due to parameter degeneracies.

\begin{figure}
\begin{center}
\includegraphics[width=\textwidth,angle=0]{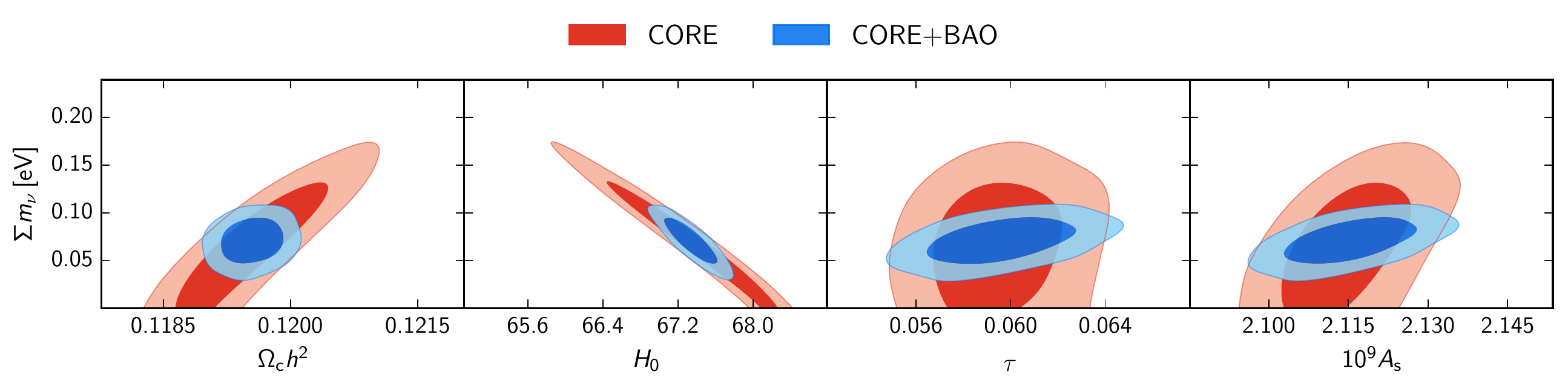}
\end{center}
\caption{
Two-dimensional marginalised constraints (68\,\% and 95\,\%)
in $\Lambda$CDM models with massive neutrinos for
\mission\ (red) and the combination of \mission\ and future BAO
measurements from DESI and \Euclid\ (blue). The fiducial model has the
minimal masses in the normal ordering with a summed mass $\sum m_\nu =
60\,{\rm meV}$.
}
\label{fig:meff}
\end{figure}

Figure~\ref{fig:meff} shows forecasted parameter constraints from
\mission\ combining the temperature and polarization power spectrum
measurements with the lensing potential power spectrum obtained from
the minimum-variance quadratic estimator. The fiducial model is close
to the minimal-mass in the normal ordering, with $\sum m_\nu =
60\,\text{meV}$, and the analysis is performed assuming degenerate
masses.\footnote{Assuming degenerate neutrinos at such low masses is clearly
inconsistent with the mass splittings inferred from neutrino
oscillations. However, cosmological observations have little
sensitivity to the mass splittings and so the constraints on the
summed mass are very similar irrespective of whether degenerate masses
or masses with realistic splittings are assumed; see Ref.~\cite{DiValentino:2016foa}
for an explicit demonstration in the context of the \mission\ mission.}
Combining
the anisotropy and lensing power spectra of \coremfive, we forecast a 1\,$\sigma$
error of $44\,\mathrm{meV}$ for the summed mass. This is a significant
improvement over current constraints, but falls someway short of the
minimum masses inferred from neutrino oscillations.

The constraint on the summed mass from \mission\ alone is limited by
degeneracies with other parameters, as shown in
Fig.~\ref{fig:meff}. The degeneracy with the Hubble constant arises
from the geometric degeneracy in the primary anisotropies. The
degeneracy with the physical density in CDM, $\Omega_{\rm c}h^2$,
arises from lensing: an increase in $\Omega_{\rm c}h^2$ pushes matter--radiation
equality to higher redshift, boosting the late-time matter power spectrum as
structure has had longer to grow in the matter-dominated
era~\cite{Pan:2014xua,Ade:2015zua}. An increase in $\Omega_{\rm c}h^2$ can therefore be
offset with an increase in the neutrino mass to preserve the lensing
power. Finally, an increase in the amplitude $A_{\rm s}$ of the primordial power
spectrum increases the lensing power spectrum proportionately on all
scales, and so is also positively correlated with the neutrino mass.

The constraint on neutrino mass can be significantly improved by
combining with measurements of the BAO feature -- a purely geometric
measurement -- in the clustering of galaxies, since these can break the
degeneracy between $\Omega_{\rm  c}h^2$ and $\sum m_\nu$. 
Increasing $\Omega_{\rm
  c}h^2$ and $\sum m_\nu$ at fixed angular scale of the CMB acoustic
peaks leads to an increase in the radial BAO observable $H(z) r_{\rm
  s}(z_{\rm drag})$ at $z >1$ and a decrease
at lower redshift, and an increase in the
angular observable $d_A(z)/r_{\rm s}(z_{\rm drag})$ (see,
e.g., Ref.~\cite{Zhen:2015yba}). Here,
$r_{\text s}(z_{\rm drag})$ is the sound horizon at the drag epoch and
$d_A(z)$ is the angular-diameter distance to redshift
$z$. Figure~\ref{fig:meff} forecasts the effect of combining \mission\
data with BAO data from DESI and \Euclid\ in the redshift range
$0.15\le z \le 2.05$, using predictions from Ref.~\cite{Font-Ribera:2013rwa} for the
BAO measurement errors. This combination 
could shrink the error on $\sum m_\nu$ to $17\,\mathrm{meV}$, giving a
high chance of a significant detection (greater than $3\,\sigma$) of
non-zero neutrino mass even for the minimal-allowed
mass.\footnote{This constraint is a little better than that reported
  in Ref.~\cite{DiValentino:2016foa} due to our inclusion of \Euclid\ BAO data, which
  helps particularly at higher redshifts ($z > 0.9$).}
Furthermore, if the total mass is close
to this minimum (around $60\,{\rm meV}$), \coremfive+BAO will likely
disfavour any total mass allowed by the inverted ordering at greater
than $2\,\sigma$ significance, providing important information on the
mass orderings.

Neutrino mass determination from CMB lensing relies on comparing the
clustering power at low redshift, determined from lensing, with the
power at last scattering, determined from the CMB
anisotropies. However, scattering at reionization reduces the \emph{observed}
anisotropy on scales smaller than the projection of the horizon size
there by a factor $e^{-\tau}$, where $\tau$ is the optical depth to
reionization. It follows that only the combination $A_{\rm s}
e^{-2\tau}$ is measured very precisely from the CMB temperature and
polarization power spectra on these scales: the $1\,\sigma$ error from \Planck\ is
$0.7\,\%$~\cite{Ade:2015xua} and we forecast $0.2\,\%$ for \mission. To separate out
$A_{\rm s}$ requires an independent measurement of the optical
depth. This can be obtained from the $E$-mode polarization data at
low multipoles, where scattering at reionization generates power
giving rise to the characteristic feature in the $E$-mode power
spectrum at $l<10$ (see Fig.~\ref{fig:noisepower}). Measuring
polarization on such large scales requires a nearly full-sky survey,
stable observations over wide separations, and excellent rejection of
Galactic foreground emission. To date, such measurements have only
been achieved from space (although efforts are underway with the
ground-based experiment CLASS~\cite{2014SPIE.9153E..1IE}). Recent results from
\planck\ give $\tau = 0.055\pm 0.009$~\cite{Aghanim:2016yuo}, while for \mission\ we forecast
a $1\,\sigma$ error of $0.002$ equal to the cosmic-variance
limit. This precision on $\tau$ limits that on $A_{\rm s}$ to around
$0.4\,\%$, and our ability to predict the lensing power spectrum for a
given mass is similarly uncertain. If the $S/N$ on a measurement of
the amplitude of the lensing power spectrum significantly exceeds
$A_{\rm s}/\sigma(A_{\rm s})$, the uncertainty in the neutrino mass
determination will be dominated by that in $A_{\rm s}$ if precision
BAO data is used to break the degeneracy with $\Omega_{\rm c}h^2$.
For \mission,
with $\sigma(\tau) = 0.002$, this corresponds to $N_{\rm modes}
\approx 1\times 10^5$, similar to what is achieved in the baseline
configuration. It follows that further improvement in the lensing
$S/N$ (i.e., increasing the sensitivity or resolution) would not lead
to proportional improvement in the measurement of neutrino mass; see
Ref.~\cite{DiValentino:2016foa} for explicit comparisons of possible design choices for \mission.

To illustrate the importance of precise determination of the optical
depth to reionization for neutrino mass constraints, we consider
replacing the large-angle polarization data from \mission\ with a
\Planck-like prior with $\sigma(\tau) = 0.01$. In this case, the
error on the summed neutrino mass
from \mission+BAO almost doubles to $30\,{\rm meV}$. This situation is
similar to that which CMB-S4 will face in the absence of a contemporaneous
space mission if attempts to measure polarization on very large scales
from the ground are unsuccessful.

\subsection{Sterile neutrinos and other massive additional relic particles}

\begin{figure}
\begin{center}
\includegraphics[width=12cm,angle=0]{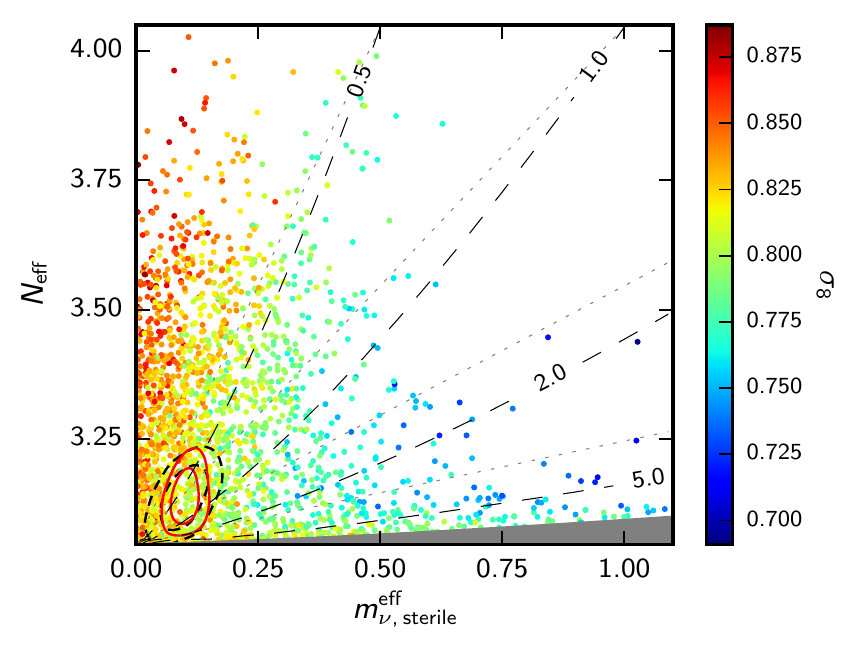}
\end{center}
\caption{
Samples from the current \textit{Planck} temperature and low-$l$
polarization data combined with BAO data (following Ref.~\cite{Ade:2015xua}) in
the  $\nnu$--$\meffsterile$ plane, colour-coded by $\sigma_8$. The models
have one massive sterile neutrino family, with effective mass $\meffsterile$,
in addition to the three active neutrinos. Dashed contours show forecast $68\,\%$ and $95\,\%$ constraints from \mission,
 and solid contours the forecast when combining with future BAO data
 from DESI and \emph{Euclid}.
The physical mass
of the sterile neutrino in the thermal scenario,
$m_{\nu,{\rm sterile}}^{\rm thermal}$, is constant along the grey dashed lines, with
the indicated mass in $\mathrm{eV}$; the grey region shows the region
excluded by the prior $m_{\nu,{\rm sterile}}^{\rm thermal}< 10\, {\rm eV}$,
which excludes most of the region where the neutrinos behave nearly like dark matter.
}
\label{fig:sterile}
\end{figure}

In addition to sterile neutrinos, many extensions to the standard
model could also produce additional relic particles, for example
thermal or non-thermal distributions of axions or gauge bosons. If
they remain relativistic until today, the main effect in the CMB is via the
increased expansion rate and anisotropic stress in the early
universe~\cite{Bashinsky:2003tk}. The former reduces power in the
damping tail at fixed angular separation of the acoustic peaks, while the anisotropic
stress introduces a characteristic phase shift in the acoustic
oscillations and hence peak locations. The contribution of non-photonic
relativistic particles to the energy density in the early universe is
usually parameterised by $N_{\rm eff}$, such that
\begin{equation}
\Delta \rho = \frac{7}{8}\left(\frac{4}{11}\right)^{4/3} N_{\rm eff}
\rho_\gamma \, ,
\end{equation}
where $\rho_\gamma$ is the energy density of photons. With this
parameterisation, the three families of active neutrinos contribute
$N_{\rm eff}  = 3.046$ and one additional sterile neutrino with the same thermal
distribution function as the active neutrinos would contribute a further $\Delta N_{\rm eff}
\approx 1$. The damping tails in the temperature and, particularly,
polarization power spectra\footnote{%
The accuracy of 
parameter inferences from the temperature power spectrum measured by 
\emph{Planck}~\cite{Ade:2015xua} are now close to being limited by errors in 
the modelling of extragalactic foregrounds. Fortunately, further progress can be 
made with the polarization anisotropies on small angular 
scales~\cite{Galli:2014kla}, since the degree of polarization of the anisotropies 
is relatively larger there (greater than 15\,\% by $l=2000$) than the foreground emission.
}
measured with
\mission\ alone gives a forecast error of $\sigma(\nnu) \approx
0.04$~\cite{DiValentino:2016foa}.

If the relic particles are massive, but are not so
massive that they look like cold dark matter in the CMB and lensing (i.e.,
physical mass less than around $10\, {\rm eV}$), \mission\ can
constrain both the mass and their contribution to $\nnu$.
As shown in
Fig.~\ref{fig:sterile}, \mission\ could dramatically reduce the allowed
parameter space compared to current \Planck\ constraints. For
detectable additional species, we forecast $\sigma(\nnu)\approx 0.04$ as for
light relics, and a $1\,\sigma$ constraint on $\meffsterile\equiv
(94.1 \Omega_{\nu,\, {\rm sterile}}h^2)\, {\rm eV}$
of  approximately $0.03\, {\rm eV}$ (or $0.02\, {\rm eV}$ including
BAO).  Here, $\Omega_{\nu,\, {\rm sterile}}h^2$ is the energy density
of the relic today and is proportional to the product of the physical
mass and $(\Delta \nnu)^{3/4}$ for a thermal relic that is now
non-relativistic.
These constraints are forecast assuming a thermal relic,
but \mission\ would give similar constraints on a variety of more general non-thermal models.  The forecast error of $\sigma(\nnu)\approx 0.04$ would be sufficient to detect at high significance any thermal relics produced after the QCD phase transition (which are currently weakly disfavoured), and is also sufficient to detect some scenarios where multiple new particles decoupled from the standard model at energies above $1\, {\rm TeV}$.

\section{Combining \mission\ lensing with other probes of
  clustering}
\label{sec:LSS}

Lensing of the CMB probes the large-scale distribution of matter in
all of the observable universe. The same structures at lower redshift
that are traced by other cosmological observables, such as the
distribution of galaxies and the coherent distortion of the shapes of
galaxies by weak gravitational lensing (cosmic shear), also lens the CMB resulting in
non-zero correlations between CMB lensing and the
tracer. Cross-correlating CMB lensing with large-scale structure
tracers is highly complementary to the auto-correlations of each
observable. Cross-correlations tend to be more robust, since they
are immune to additive systematic effects that are independent between the
observables. Moreover, cross-correlating allows calibration of
multiplicative effects, such as galaxy bias or multiplicative bias in
the estimation of galaxy shapes, which would otherwise compromise the
cosmological information that can be extracted from the observable.

\mission\ will produce a high-$S/N$ lensing map over nearly the full
sky, allowing a wealth of cross-correlation science with current and
future large-scale structure surveys. In this section, we
highlight the potential for cross-correlating CMB lensing from
\mission\ with two particularly important tracers: lensing of galaxies
and galaxy clustering. We also summarise areas where cross-correlation
of lensing and other fields may advance our understanding
of astrophysics at high redshift.

\subsection{Galaxy lensing}

Lensing by large-scale structure can be probed in optical imaging
surveys through its shearing effect on the shapes of background galaxies.
Galaxy lensing is a key observable of ongoing (e.g., DES~\cite{Becker:2015ilr} and KiDS~\cite{Hildebrandt:2016iqg})
and future imaging surveys (e.g.,
LSST~\cite{2012arXiv1211.0310L} and
\emph{Euclid}~\cite{2013LRR....16....6A}). With approximate redshifts
for the source galaxies, it is possible to map the evolution of cosmic
shear over time (tomography) and so probe the growth of structure and the cosmic
expansion history and hence the physics of cosmic acceleration (see below).

CMB lensing is highly complementary to galaxy lensing. Although the CMB reconstruction is at lower resolution, and
lacks the tomographic aspect accessible with galaxy lensing, it probes
higher redshifts, and the $S/N$ is dominated by clustering in the
well-understood linear regime. By constrast, most of the potential
$S/N$ for galaxy lensing is deep in the non-linear regime where
modelling uncertainties are larger. Generally, CMB and galaxy lensing
are affected by very different systematic effects. For the latter,
 intrinsic alignments in the shapes of galaxies due to the local tidal
environment in which they form (see Ref.~\cite{Joachimi:2015mma} for a review), source
redshift errors, and biases in the estimation of the shapes of
galaxies are all important. In practice, the combination of CMB and
galaxy lensing with overlapping footprints on the sky is particularly
promising. For example, their cross-correlation allows self-calibration of
multiplicative biases in the galaxy shape
measurements~\cite{2012ApJ...759...32V,Das:2013aia,Schaan:2016ois} and
models for the intrinsic-alignment
signal~\cite{Hall:2014nja,Troxel:2014kza,2016MNRAS.461.4343L}. The
correlation between CMB and galaxy lensing has been detected recently
at modest significance using a range of
surveys~\cite{Hand:2013xua,Liu:2015xfa,Kirk:2015dpw,Singh:2016xey,Harnois-Deraps:2017kfd}. With
\mission\ and, for example, \Euclid\ lensing, the amplitude of
the total cross-correlation will be measured with a $S/N$ of around
170.
The combination of CMB and galaxy lensing will also yield parameter constraints
that are more robust against degeneracies with other parameters. We now
illustrate some of these ideas in the context of constraints on neutrino mass and dark energy.

\paragraph{Absolute neutrino masses}

The constraints on the absolute mass scale of neutrinos from \mission\
(Sec.~\ref{sec:mnu}) are comparable to those forecast for
other future probes of clustering, including cosmic shear
measurements from \Euclid~\cite{2013LRR....16....6A}.
Even stronger and, importantly, more robust neutrino mass constraints can be
obtained by combining \mission\ with such probes. 
As an
illustration of the robustness against parameter degeneracies,
a conservative forecast\footnote{We include wavenumbers only up to
  $k_\mathrm{max}=0.5h\,{\rm Mpc}^{-1}$ in the analysis, to avoid
  systematic uncertainties associated with non-linear clustering.}
for the combination of \mission, BAO, and \Euclid\ cosmic shear in models
with spatial curvature gives an error on the summed mass of active
neutrinos of less than $20\,{\rm meV}$ (from $16\,{\rm meV}$ without free curvature),
so at least a $3\,\sigma$ detection of non-zero mass is still likely~\cite{DiValentino:2016foa}.
In contrast, with current CMB data the degradation would be much worse: 
for the combination \Planck+BAO+\Euclid\ cosmic shear the
degradation in errors when marginalising over free curvature is from
$23\,{\rm meV}$ to $33\,{\rm meV}$. 

\paragraph{Dark energy and modifications to gravity}

Understanding the observed late-time accelerated expansion of the
Universe is a critical problem for fundamental physics. While current
observations are consistent with acceleration being due to a
cosmological constant (the $\Lambda$CDM model), its unnaturally small value has led to the
development of alternative theories such as those involving (dynamical)
dark energy or modifications to the laws of gravity on large
scales. Probing the underlying physics of cosmic acceleration, through
measurements of the expansion history and growth of structure,
is a key
science goal for Stage-IV dark energy experiments (e.g., DESI, LSST,
and \textit{Euclid}). The effects of dark energy are degenerate in the primary CMB
fluctuations, which originate at much higher redshift than the onset
of cosmic acceleration ($z \approx 1$). However, through secondary
effects in the CMB, \mission\ will provide several dark energy
observables that complement other low-redshift probes: the 
cluster sample detected with \mission\ via the thermal
Sunyaev-Zel'dovich (SZ) effect~\cite{Melin:2017lkr} (see also Sec.~\ref{sec:clustermass}); peculiar velocities as
measured by the kinetic SZ effect~\cite{Melin:2017lkr}; and CMB lensing. 

Lensing of the CMB \emph{alone} is not a very powerful discriminant of
models in which dark energy is only dynamically important at late
times, since most, though not all, of the lensing effect in the CMB is sourced at too
high a redshift. However, cross-correlation with tracers of
large-scale structure at redshifts $z < 1$ isolates the lensing
contribution during the period when dark energy is significant.
As a probe of
dark energy, CMB lensing from \mission\ will therefore be particularly
powerful when combined with galaxy lensing and galaxy clustering data across
redshift. 

Combining \mission\ lensing with tomographic measurements
of galaxy lensing adds a precisely determined high-redshift source plane and,
as discussed above, allows cross-calibration of the majority of the expected galaxy lensing 
systematic effects. To illustrate these ideas, we consider constraints
on dark energy models with equation of state parametrised in terms of
the scale factor $a$ as $w(a) = w_0 + w_a (1-a)$, marginalising over
the absolute neutrino mass and galaxy lensing systematic effects
following Ref.~\cite{Kitching:2014lga}. We
present results in terms of the dark energy figure of merit,
$\text{FoM} = [\text{det}\,\text{cov}(w_0,w_a)]^{-1/2}$. With \Euclid\
cosmic shear alone, the FoM is very dependent on whether or not
poorly-understood non-linear
scales are included in the analysis, degrading by an order of
magnitude if the maximum wavenumber is reduced from
$k_\mathrm{max}=5.0h\,{\rm Mpc}^{-1}$ ($\text{FoM}\approx 50$)
to $1.5h\,{\rm Mpc}^{-1}$ ($\text{FoM}\approx 5$).
Combining with \mission\ data
helps considerably, improving the FoM to approximately 300 using only
linear scales from \Euclid. These improvements will be significantly greater ($\text{FoM} \approx 2400$)
if strategies developed for internal calibration of \Euclid\ data are successful (e.g., using image
simulations to calibrate multiplicative bias in the estimation of
galaxy shear).
In this
way, we can recover dark energy science from cosmic shear with
\Euclid\ using only relatively clean (quasi-)linear scales.

\subsection{Galaxy clustering}

Galaxies form preferentially within overdensities of the large-scale
distribution of dark matter. Galaxy clustering is therefore potentially a powerful probe of the
underlying mass distribution across cosmic time, and so of dark energy,
modifications to gravity, neutrino masses, and the statistics of the
primordial perturbations. Forthcoming galaxy redshift surveys (such as
DESI, \Euclid, and LSST) will extend significantly the statistical
power of galaxy clustering measurements due to their large survey
volumes, depths, and accuracy of redshifts. 

A key issue in the interpretation of galaxy surveys is the uncertain
relation between the clustering of galaxies and dark matter. 
On large
scales, this is generally parameterised by a bias function $b(z)$,
which depends on redshift as well as galaxy properties. Uncertainty in
the bias limits the cosmological information that can be extracted
from the broadband galaxy power spectrum. 
Lensing helps significantly
in this regard since it probes the clustering of all mass along the
line of sight back to the source. 
Lensing of background
sources is correlated with the clustering of foreground galaxies, as the
same large-scale structures that are traced by the foreground
galaxies lens the background sources. By comparing the
cross-correlation between the lensing of
background sources and the galaxy overdensity (within some redshift
range centred on $z$) with the auto-power spectrum of the galaxy
overdensity, one can separate the bias $b(z)$ and clustering amplitude
at that redshift with only weak model
dependencies. For high-redshift galaxies, CMB lensing is particularly
helpful as the last-scattering surface is so distant. This approach
has recently been demonstrated with galaxies from
DES~\cite{Giannantonio:2015ahz} and, at higher redshift, from
\textit{Herschel}~\cite{Bianchini:2015yly}. These tomographic analyses
follow earlier work using projected galaxy
samples~\cite{Smith:2007rg,Hirata:2008cb,Feng2012,Bleem:2012gm,Sherwin:2012mr,Bianchini:2014dla,Kuntz:2015wza,Omori:2015qda}. With
forthcoming galaxy clustering data, and high-$S/N$ CMB lensing
measurements over large fractions of the sky, this tomographic
approach will allow precise tests of the growth of structure
complementing other probes such as tomographic cosmic shear,
redshift-space distortions, and the number counts of galaxy clusters.




\subsection{High-redshift astrophysics}

More generally, cross-correlating CMB lensing with other probes of
large-scale structure has great promise as a probe of astrophysics at
high redshift. A recent highlight of this approach is constraining
the high-redshift star formation rate from correlations between CMB
lensing and clustering of the cosmic infrared
background (CIB)~\cite{Song:2002sg,Ade:2013aro,Holder:2013hqu,vanEngelen:2014zlh,Ade:2013hjl}
-- the unresolved flux from dusty, star-forming galaxies. In contrast
to the CIB spectra across frequencies, the
cross-correlation with lensing is insensitive to residual Galactic dust emission in the CIB
maps, and does not require separation of the shot noise that arises
from Poisson fluctuations in the number density of the galaxies that
contribute to the CIB. A further application is
constraints on gas physics in low-mass clusters and groups of
galaxies, gas that is otherwise difficult to detect,
from the correlation between CMB lensing and maps of the diffuse
thermal SZ effect\footnote{The thermal
  SZ effect (e.g., Ref.~\cite{Sunyaev:1980vz}) is the Compton scattering of the CMB
off hot ionized gas. Its characteristic frequency dependence allows
separation from other emission components in multi-frequency
maps.}~\cite{Hill:2013dxa}. Such measurements will be significantly
advanced with the diffuse tSZ map from \mission~\cite{Melin:2017lkr},
which should be much cleaner than the equivalent from
\Planck~\cite{Aghanim:2015eva}, and the improved $S/N$ of the \mission\ lensing map.
As a final application, we note the recent constraints on the bias and
hence halo masses of
high-redshift quasar hosts from cross-correlation of CMB lensing with
quasar catalogues~\cite{Sherwin:2012mr,Geach:2013zwa}. With higher
precision CMB lensing maps, such studies will be extended to probe
dependencies on the quasar properties, such as redshift and
luminosity.

\section{Delensing $B$ modes}
\label{sec:delens}

One of the main science goals of \mission\ is to search for the
distinctive signature of primordial gravitational waves in $B$-mode
polarization~\cite{Kamionkowski:1996zd,Seljak:1996gy}.
Primordial gravitational waves are a critical test of
cosmic inflation in the early universe, and their detection would
determine the energy scale at which inflation occurred and provide
important clues to the physics of inflation. The inflationary science
case for \mission\ is discussed in detail in Ref.~\cite{Finelli:2016cyd}.

Lensing of the CMB
converts $E$-mode polarization into $B$-mode~\cite{Zaldarriaga:1998ar}, and these lens-induced
$B$-modes are a source of confusion in searches for primordial
gravitational waves. However, it is possible partially to remove the
lensing $B$-modes in a process known as ``delensing''; essentially,
this involves remapping the observed polarization with an estimate of
the CMB lensing
deflections~\cite{Kesden:2002ku,Knox:2002pe,Seljak:2003pn}. In this
section, we discuss the prospects for delensing with \mission.

\begin{figure}
\begin{center}
\includegraphics[width=8cm,angle=-90]{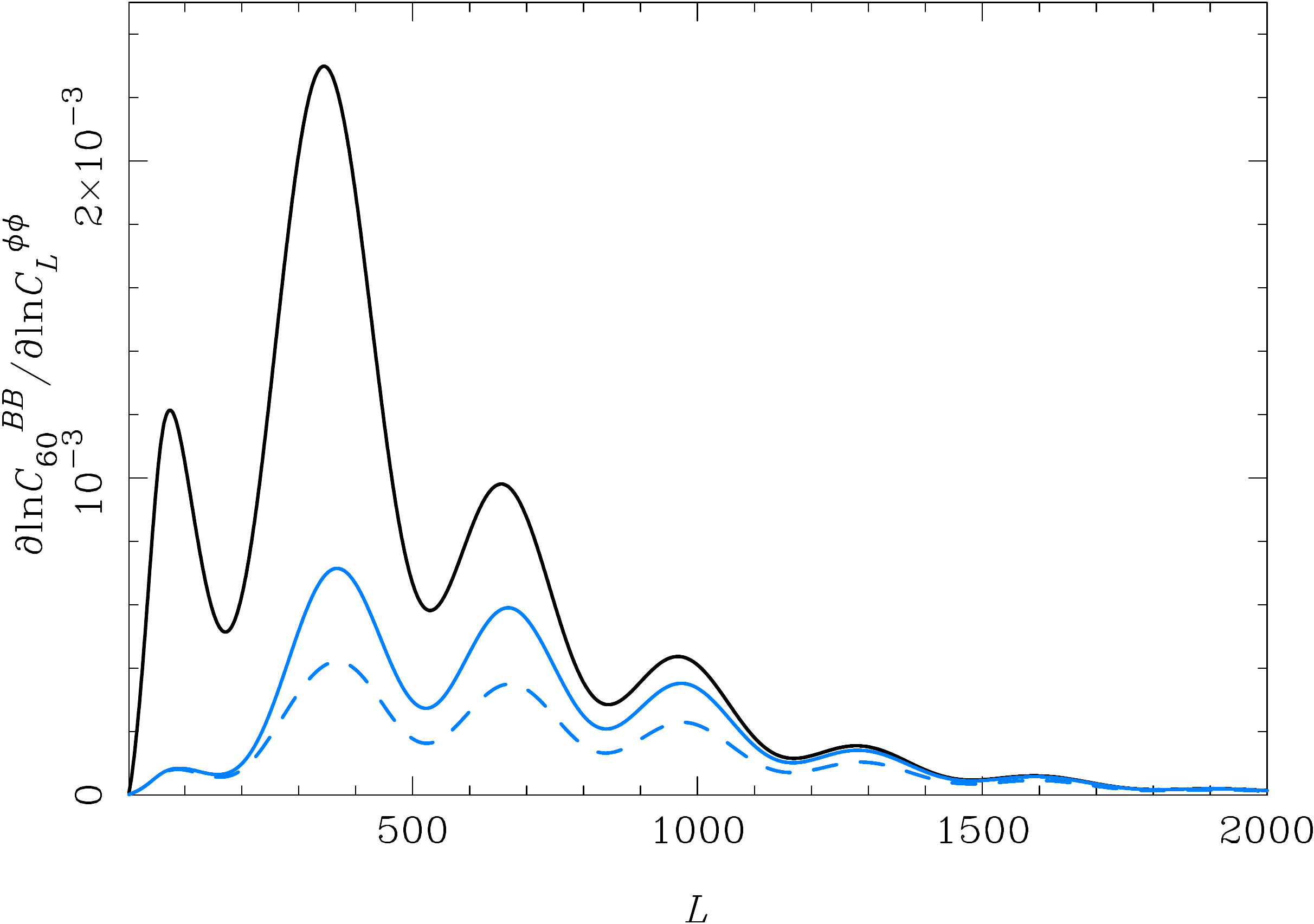}
\end{center}
\caption{Fractional contribution to the lens-induced $B$-mode power at
multipole $l=60$ per multipole of the lensing potential. The black
line shows the contribution before delensing, and so the area under
the curve is unity. The impact of delensing is to suppress the
contributions from lenses on scales where the $S/N$ on the
reconstructed lensing potential is high. This suppression is shown for
internal delensing with \mission\ (solid blue), in which case the
lensing power is reduced by 60\,\%, corresponding to a reduction in the
error on $r$ (for $r=0$) by a factor $1.9$. Combining with measurements of the
CIB from \mission\ further helps suppress the smaller-scale lenses
where the $S/N$ on the reconstructed lensing potential is larger (blue
dashed). In this case, the lensing power is reduced by 70\,\%,
corresponding to a reduction in the error on $r$ of $2.5$.}
\label{fig:BBlensingcontribs}
\end{figure}

The $B$-modes produced from conversion of $E$-mode polarization by
lensing are approximately
\begin{equation}
B_{lm}^{\text{lens}} = -i(-1)^m\sum_{LM}\sum_{l'm'}\left( \begin{array}{ccc} l & L & l'
                                          \\ -m & M &
                                                      m' \end{array}\right)
{}_- F^2_{l L l'} \phi_{LM} E_{l'm'} \, ,
\label{eq:BfromE}
\end{equation}
where the geometric coupling term ${}_- F^2_{l L l'}$ is given in
Appendix~\ref{app:quadrecon}. The power spectrum $C_l^{BB,\text{lens}}$ is
therefore
\begin{equation}
C_l^{BB,\text{lens}} \approx \frac{1}{2l+1}\sum_{Ll'} \left({}_- F^2_{lLl'}\right)^2
C_L^{\phi\phi} C_{l'}^{EE} \, ,
\label{eq:sphClBB}
\end{equation}
and is shown in the left-hand panel of Fig.~\ref{fig:noisepower}. For multipoles
$l\lesssim 400$, $C_l^{BB,\text{lens}} \approx 2.0 \times 10^{-6} \,\mu\text{K}^2$ is almost constant, and so
lens-induced $B$-modes act like an additional $5\,\mu{\rm K}\,{\rm
  arcmin}$ of white noise on all scales relevant for
detection of $B$-modes from primordial gravitational waves. This
behaviour follows from the low-$l$ limit of Eq.~\eqref{eq:sphClBB},
which gives
\begin{equation}
C_l^{BB,\text{lens}} \approx \frac{1}{2}\sum_{l'} \frac{2l'+1}{4\pi}
l'{}^4 C_{l'}^{\phi\phi} C_{l'}^{EE} \, .
\label{eq:lowlClBB}
\end{equation}
At multipoles $l>10$, the $B$-mode lensing power spectrum exceeds that
from primordial gravitational waves if the tensor-to-scalar
ratio\footnote{The tensor-to-scalar ratio is the ratio of the
  primordial power spectra of gravitational waves and curvature
  fluctuations at a pivot scale $k_\ast$. Here, we adopt $k_\ast =
  0.05\,\text{Mpc}^{-1}$.} $r \gtrsim 0.01$ (see
Fig.~\ref{fig:noisepower}). The best limits on $r$ now come from
$B$-mode polarization, with the combination of BICEP/Keck Array data
and \planck\ and \WMAP\ data (primarily to remove foreground emission from our
Galaxy) giving $r<0.09$ at 95\,\% C.L.~\cite{Array:2015xqh}.
Large-scale lensing $B$-modes are produced
from $E$ modes and lenses over a broad range of
scales, with 50\,\% of the power at a multipole of $60$ coming from lenses
at multipoles $L>400$. This is illustrated in
Fig.~\ref{fig:BBlensingcontribs}, where we plot the 
fractional contribution to the lens-induced $B$-mode power at a
multipole of $60$ per multipole of the lensing potential, i.e., $d \ln
C_{60}^{BB,\text{lens}}/d \ln C_L^{\phi\phi}$. The generation of
large-scale $B$-modes from $E$-modes and lenses on significantly
smaller scales is the origin of the white-noise behaviour of $C_l^{BB,\text{lens}}$.

The lensing $B$-mode power spectra can be accurately predicted in any
model\footnote{Non-linear corrections to the matter power spectrum
  contribute to $C_l^{BB,\text{lens}}$ at around the 6\,\%
  level on large scales~\cite{Challinor:2005jy}. The impact of systematic uncertainties in modelling the
  small-scale matter power spectrum, including the effects of baryonic
  physics, is a small change in the amplitude of
  $C_l^{BB,\text{lens}}$ on large scales. This can be dealt with by
  marginalising over the amplitude of $C_l^{BB,\text{lens}}$ during
  parameter estimation.}, with the uncertainty due to parameter errors
at around the $0.5\,\%$ level for \mission. The main impact of lensing
on the estimation of the primordial gravitational wave amplitude is
therefore not from the average power that lensing contributes,
subtraction of which causes only a small increase in parameter uncertainties,
but rather the increased sample variance. We can illustrate the issue
with the following crude approximation to the error on the tensor-to-scalar
ratio estimated from the $B$-mode power spectrum:
\begin{equation}
\frac{1}{\sigma^2(r)} \sim f_{\rm sky} \sum_l
\frac{2l+1}{2}\left(\frac{C_l^{BB,\text{gw}}(r=1)}{r
    C_l^{BB,\text{gw}}(r=1) + C_l^{BB,\text{lens}} + N_l^{BB}}
\right)^2 \, ,
\label{eq:knoxBB}
\end{equation}
where $C_l^{BB,\text{gw}}(r=1)$ is the $B$-mode power spectrum from
primordial gravitational waves for $r=1$ and $N_l^{BB}$ is the power
spectrum of the instrument noise. The presence of
$C_l^{BB,\text{lens}}$ on the right-hand side describes the effect of
the sample variance of the lens-induced $B$-modes. This becomes
important as noise levels approach $5\,\mu{\rm K}\,{\rm arcmin}$, and
for an experiment such as \mission\ is the dominant source of
``noise''. Indeed, ground-based experiments have already reached this
sensitivity for observations covering a few hundred square degrees
(before foreground cleaning)~\cite{Array:2015xqh}. 

\begin{figure}
\begin{center}
\includegraphics[width=5.25cm,angle=-90]{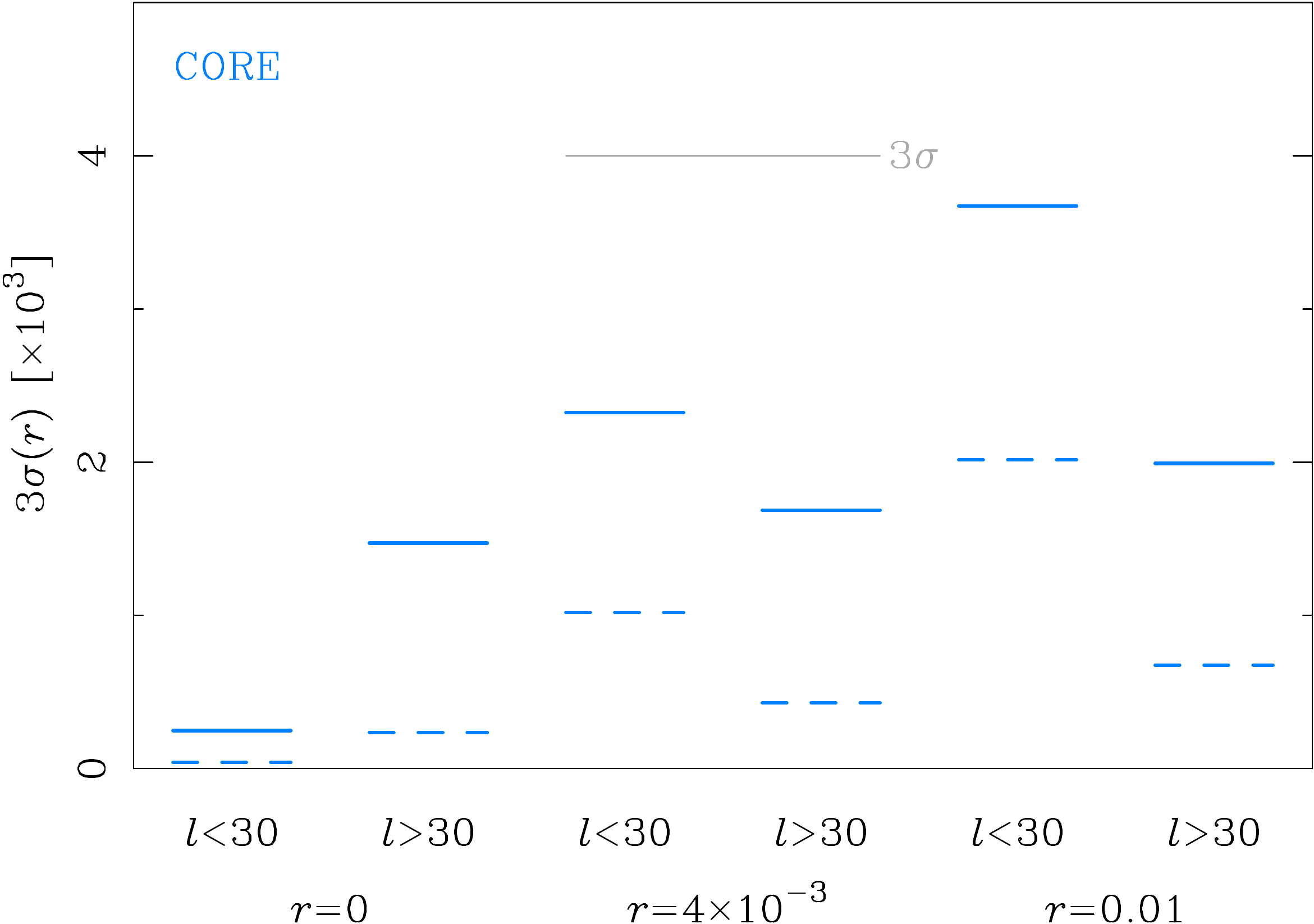}
\includegraphics[width=5.25cm,angle=-90]{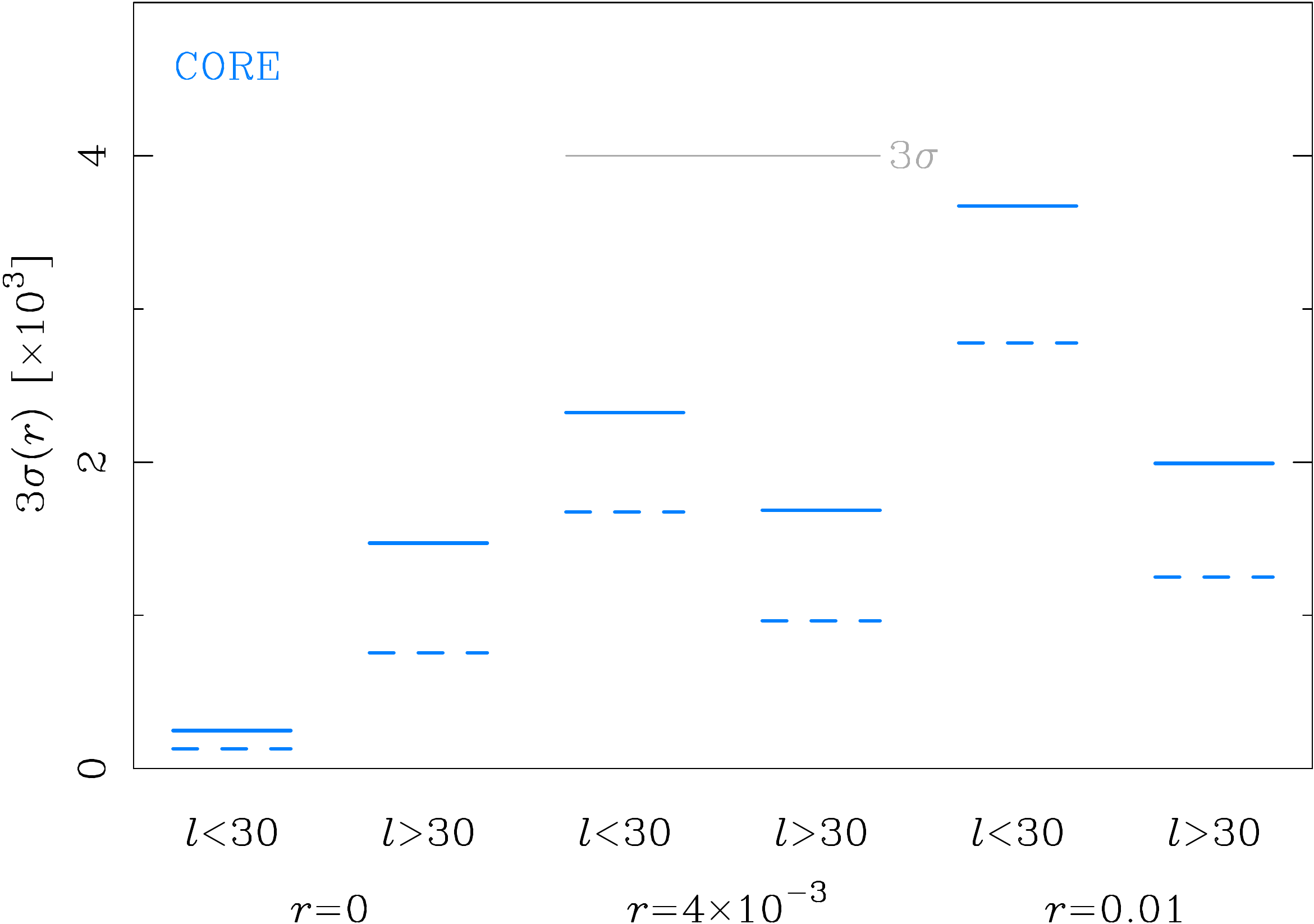}
\end{center}
\caption{Impact of lensing on constraints on $r$ from \mission\ for models with
  $r=0$, $r=4\times 10^{-3}$ (typical for the Starobinsky model), and
  $r=0.01$. For each model, the $3\,\sigma$ error (with all
  other parameters fixed) is shown using only $BB$ for multipoles
  $l<30$ (i.e., the signal from reionization) and $l>30$ (the signal
  from recombination) over $70\,\%$ of the sky. The grey line shows the
  $3\,\sigma$ threshold for detecting $r=4\times 10^{-3}$. In the left-hand plot, the solid lines assume no delensing, while the
  dashed lines assume perfect delensing. In the
  right-hand plot, the solid lines again assume no delensing, but the
  dashed lines assume internal delensing. Effects of foreground
  removal are not included other than through our use of the weighted
  combination of the six 130--220\,GHz channels for the
  effective instrument noise power.}
\label{fig:lensingimpactonr}
\end{figure}

The impact of lensing sample
variance is shown in Fig.~\ref{fig:lensingimpactonr}. The $3\,\sigma$
error on $r$ is shown, based on Eq.~\eqref{eq:knoxBB}, for three
models: (i) $r=0$; (ii) $r=4 \times 10^{-3}$, typical of
the $R^2$ Starobinsky model~\cite{Starobinsky:1980te} that predicts $r=12/N_*^2$,
where $N_* \approx 55$ is the number of $e$-folds between the end of inflation and
the time that modes of wavenumber $k_*$ exited the Hubble radius
during inflation; and (iii) $r = 0.01$, roughly the forecasted
detection limit of the current generation of sub-orbital
experiments. Starobinsky inflation is an example of a model with a red
spectrum of curvature fluctuations, with a tilt $n_s - 1 \propto -1/N_\ast$,
which fits the measured temperature and $E$-mode polarization power
spectra but produces a small $r \propto 1/N_\ast^2$. Such models will
be natural targets for \mission\ if large-field models with $r \propto
1/N_\ast$ are ruled out by the time of flight. The errors on $r$ are
shown based on $B$-modes with $l<30$ and $l>30$. The former is
intended to emphasise the constraints arising from the signal
generated at reionization (see Fig.~\ref{fig:noisepower}), while the latter
isolates the signal from scattering around recombination. As discussed in Sec.~\ref{sec:mnu},
measuring the signals from reionization is likely only possible from space,
but will be very important to confirm that any $B$-mode signal detected on
degree scales is indeed due to primordial gravitational waves. We can
draw the following conclusions from Fig.~\ref{fig:lensingimpactonr}.
\begin{itemize}
\item In the limit that the signal sample variance is small compared
  to the lensing sample variance on all relevant scales, i.e., $r
  C_l^{BB,\text{gw}}(r=1) \ll C_l^{BB,\text{lens}}$, lensing increases
  the error on $r$ by $1+C_l^{BB,\text{lens}}/N_l^{BB} \approx 6.5$
  (for \mission)
  from both the reionization and recombination signals.
\item For larger $r$, lensing has relatively more of an impact on the
  recombination signal than the reionization signal since
  $C_l^{BB,\text{gw}} / C_l^{BB,\text{lens}}$ is boosted at $l<10$ by
  reionization.
\item Lensing would limit the 
ability to test models such as Starobinsky inflation at very high
significance on both reionization and recombination scales.
For example, for $r=4\times
10^{-3}$, the sample
variance of the lens-induced $B$-modes would limit the $S/N$ with \coremfive\
to $7.1$ from the recombination signal ($l>30$), and $S/N=5.1$ from the
reionization signal at low multipoles. This situation is worsened
for observations over smaller sky fractions.
\end{itemize}

To reduce the impact of sample variance of the lens-induced $B$-modes
requires their coherent subtraction. Fortunately, such delensing is
possible by combining the precise measurements of $E$-mode
polarization from \coremfive\ with its lensing
reconstruction. There are several ways to implement delensing, but for large-scale
$B$-modes, where the gradient approximation of Eq.~\eqref{eq:BfromE} is
accurate, subtraction of a template constructed from the
Wiener-filtered lens reconstruction and the Wiener-filtered
$E$-mode polarization is close to optimal:
\begin{equation}
\hat{B}_{lm}^{\text{lens}}  = -i(-1)^m\sum_{LM}\sum_{l'm'}\left( \begin{array}{ccc} l & L & l'
                                          \\ -m & M &
                                                      m' \end{array}\right)
{}_- F^2_{l L l'} \mathcal{W}_L^\phi \hat{\phi}_{LM}
\mathcal{W}_{l'}^E E^{\text{dat}}_{l'm'} \, .
\label{eq:Btemplate}
\end{equation}
Here, the Wiener filters are $\mathcal{W}^\phi_l = C_l^{\phi\phi} /
(C_l^{\phi\phi} + N_l^{(0)})$ and $\mathcal{W}^E_l =
C_l^{EE}/(C_l^{EE} + N_l^{EE})$, $\hat{\phi}$ is the lens
reconstruction, and $E^{\text{dat}}$ is the observed (noisy) $E$-mode
polarization after deconvolution of the instrument beam.
After subtracting the synthetic $B$-modes in Eq.~\eqref{eq:Btemplate}
from the observed $B$-modes, the residual lensing power is
approximately
\begin{equation}
C_l^{BB,\text{delens}} \approx \frac{1}{2l+1}\sum_{Ll'} \left({}_- F^2_{lLl'}\right)^2
C_L^{\phi\phi} C_{l'}^{EE} \left(1-\mathcal{W}_{L}^{\phi}
  \mathcal{W}_{l'}^{E}\right) \, .
\label{eq:BBresidual}
\end{equation}
In the limit that the $S/N$ on the $E$-mode polarization is large on
the scales relevant for lensing conversion to large-angle $B$-mode
polarization, $\mathcal{W}^E_l \approx 1$. The contribution to the
residual $B$-mode power from
lenses at multipole $L$ is therefore suppressed by a factor
$1-\mathcal{W}_L^{\phi}$, so that $\mathcal{W}_L^{\phi}$ gives the scale-dependent delensing
efficiency. 
Figure~\ref{fig:BBlensingcontribs} shows the contribution
to the residual $B$-mode power per lensing multipole as a fraction of
the original lensing power at multipole $l=60$, i.e.,
$(1-\mathcal{W}_L^{\phi}) d \ln C_{60}^{BB,\text{lens}}/d \ln
C_L^{\phi\phi}$, for the minimum-variance lens reconstruction with
\mission. The contribution is strongly suppressed for lenses on scales
where the $S/N$ on the reconstruction is high, making the delensed
spectrum even closer to white noise on large scales than the spectrum
before delensing.
The integrated effect
is a reduction of 60\,\% in $C_l^{BB,\text{lens}}$ by internal
delensing. The impact for constraints on $r$ with \mission\ is
illustrated in the right-hand plot in
Fig.~\ref{fig:lensingimpactonr}.
Here, we have assumed that the residual $B$-modes after delensing are
approximately Gaussian on large scales [as we also assumed in
Eq.~\eqref{eq:knoxBB}].
For Starobinsky inflation, internal delensing improves the $S/N$ on $r$
to 12.5 from $l>30$, allowing critical tests of this important class of
models through detailed characterisation of the $B$-mode spectrum.
For models with very low $r$, delensing improves $\sigma(r)$ by a
factor of two on all scales. Internal delensing of $B$-modes (and the
temperature and $E$-mode polarization) has recently been demonstrated
with data from \Planck, although the \Planck\ reconstruction noise
means that only around 7\,\% of the $B$-mode lensing power can currently be removed~\cite{Carron:2017vfg}.

Internal lens reconstructions from the CMB are noisy on small scales
that still contribute significantly to the large-angle $B$-mode
power. The inclusion of other tracers of the lensing potential with
better $S/N$ on small scales can therefore further improve $B$-mode
delensing. The cosmic infrared background (CIB) is a particularly
promising tracer~\cite{Simard:2014aqa,Sherwin:2015baa}, since it
is highly correlated (around 80\,\%) with CMB
lensing~\cite{Ade:2013aro}. In principle, delensing with the CIB alone
can remove around 60\,\% of the lensing $B$-mode power but this
requires very accurate subtraction of Galactic dust emission (in total
intensity) when estimating the CIB from multi-frequency data. CIB
delensing has recently been demonstrated in practice, both for
delensing temperature anisotropies~\cite{Larsen:2016wpa} and $B$-mode
polarization~\cite{Manzotti:2017net}. We can also optimally combine an internal lens
reconstruction and the CIB (see Ref.~\cite{Yu:2017djs} for a recent example
with \Planck\ maps).  The high-frequency channels of \mission\
make it uniquely capable of separating the CIB from Galactic dust.
On large scales, the optimal combination
is dominated by the lens reconstruction, while the CIB dominates on
smaller scales where the $S/N$ on the lens reconstruction is poor.
Note that on these small scales, any residual dust
contamination in the estimated CIB is less significant and a high
degree of correlation with lensing can be maintained. Generally, for
$N$ tracers, $I_i$, of the lensing potential, with (cross-)power
spectra $C_l^{I_i I_j}$ amongst themselves and $C_l^{I_i \phi}$ with
the lensing potential, the optimal combination for delensing is
\begin{equation}
\phi_{lm,\text{WF}} = \sum_{ij} C_l^{I_i\phi}
[\text{\textsf{C}}_l^{-1}]_{ij} I_{lm,j} \, ,
\end{equation}
where the components of the matrix $\text{\textsf{C}}_l$ are $C_l^{I_i
  I_j}$. Using $\phi_{lm,\text{WF}}$ to construct the $B$-mode
template~\eqref{eq:Btemplate}, the residual power after delensing is still
given by Eq.~\eqref{eq:BBresidual} but with $\mathcal{W}_l^\phi$ replaced with
$\rho_l^2$, where $\rho_l$ is the correlation coefficient between
$\phi_{\text{WF}}$ and $\phi$ with
\begin{equation}
\rho_l^2 = \frac{\sum_{ij} C_l^{I_i\phi}
[\text{\textsf{C}}_l^{-1}]_{ij} C_l^{I_j\phi}}{C_l^{\phi\phi}} \, .
\end{equation}
We show the product $(1-\rho_L^2) d \ln C_{60}^{BB,\text{lens}}/d \ln
C_L^{\phi\phi}$ in Fig.~\ref{fig:BBlensingcontribs} for the combination of the
minimum-variance lens reconstruction from \mission\ and the CIB at
500\,GHz, using models from Ref.~\cite{Ade:2013aro} for the CIB spectra. 
We assume negligible dust contamination and instrument noise in the
CIB map.
With this approach, we can remove 70\,\% of the
lensing $B$-mode power on large angular scales (cf.\ 60\,\% without the
CIB), which corresponds to an improvement in the tensor-to-scalar
ratio (for $r=0$) by a factor of $2.5$ compared to no delensing.

\begin{figure}
\begin{center}
\includegraphics[width=12cm,angle=0]{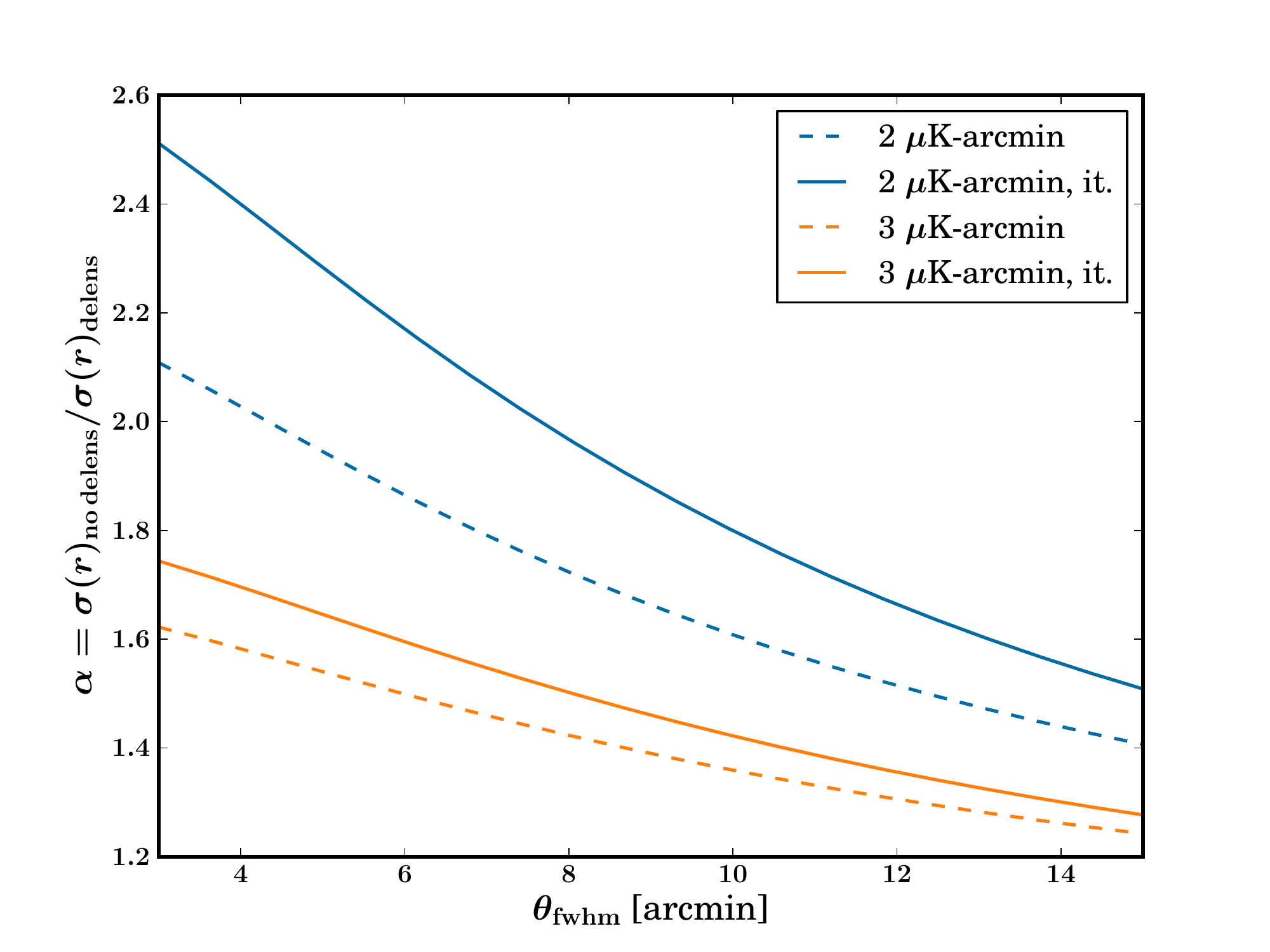}
\end{center}
\caption{Fractional improvement in constraints on $r$, assuming $r=0$,
  by internal delensing as a function of the angular resolution in the
  range relevant for space-based experiments. Results are shown for
  polarization noise levels of $3\,\mu\text{K\,arcmin}$ (orange) and
  $2\,\mu\text{K\,arcmin}$ (blue), without (dashed) and with (solid)
    iterative delensing.}
\label{fig:delensingperf}
\end{figure}

Finally, we return to the issue of more optimal lens reconstruction
that we discussed briefly in Sec.~\ref{sec:recon}. There, it was noted that
one can improve over reconstructions based on quadratic estimators for
noise levels comparable to or better than the lensing $B$-mode noise (i.e.,
$5\,\mu{\rm K}\,{\rm arcmin}$). The noise levels on reconstructions
that properly maximise the posterior distribution of $\phi$ given the observed
CMB fields have been shown (in simulations; e.g.,~Ref.~\cite{Carron:2017mqf}) to be well reproduced by an
approximate iterative calculation of the noise power $N_l^{(0)}$ of the
quadratic estimator~\cite{Smith:2008an}. Here, we use the implementation
described in
Ref.~\cite{Errard:2015cxa}, which uses only the $EB$ estimator.
Figure~\ref{fig:delensingperf} shows the
fractional improvement in $\sigma(r)$, for $r=0$, from iterative
delensing and the simple quadratic estimator compared to no
delensing. The comparisons are made as a function of angular
resolution and for two representative polarization noise levels. For
the baseline specifications of \mission\ (effective beam size of
$6.2\,\text{arcmin}$ and $2.1\,\mu{\rm K}\,{\rm arcmin}$ noise), the
improvement in $\sigma(r)$ is around $1.9$ for the quadratic estimator
and $2.2$ for iterative delensing.\footnote{The result for the
  quadratic estimator is a little worse than that quoted earlier, which was
  based on the minimum-variance quadratic estimator.} For non-zero
$r$, the relative gain in $\sigma(r)$ from iterative delensing would
be smaller. More substantial gains are achieved at higher angular
resolution and with lower noise, and so optimising delensing will be
important for forthcoming deep ground-based surveys.

\section{Cluster mass calibration}
\label{sec:clustermass}

The abundance of galaxy clusters as a function of mass and redshift is
a sensitive probe of the evolution of density fluctuations at late
times. In particular, it is sensitive to the matter density parameter,
$\Omega_{\rm m}$, the equation of state of dark energy, $w(a)$, and the
amplitude of the fluctuations, $\sigma_8$.
In recent years, large cluster samples have been assembled
with clusters detected via the thermal Sunyaev-Zel'dovich (tSZ) effect
in data from \planck~\cite{Ade:2015fva},
ACT~\cite{Hasselfield:2013wf}, and SPT~\cite{Bleem:2014iim}. Compared
to selection in other wavebands, the tSZ approach has a
particularly well-understood selection function and can be extended to
high redshifts. In its baseline configuration, \mission\ will detect
around 40\,000 clusters over the full sky ($S/N \geq 5$),
significantly extending current catalogues. The combination of the
many frequency channels of \mission\ and the deep, high-resolution imaging
that is possible from the ground (e.g., with CMB-S4) is particularly powerful and could
detect around $200\,000$ clusters~\cite{Melin:2017lkr}. The
statistical power of such catalogues is very high, but in order to
extract cosmological information from cluster abundances accurate
estimates of cluster masses are needed. Cluster masses can be
estimated via the cluster X-ray signal assuming hydrostatic
equilibrium, an assumption that can be violated in several scenarios
(e.g., bulk motions in the gas or nonthermal sources of
pressure~\cite{Nagai:2007,Piffaretti:2008,Meneghetti:2010}). Alternatively,
galaxy lensing offers another way to estimate cluster masses via the
cluster-induced gravitational shear (see, e.g.,
Ref.~\cite{Battaglia:2015zfa} in the context of tSZ-selected samples).
This approach is independent of
the complex baryonic physics involved in X-ray estimates and directly
probes the total mass. However, it is difficult to extend to
high-redshift clusters due to the paucity of background sources and
the uncertainty in source redshifts.

It has long been suggested that CMB lensing can be used to measure
cluster
masses~\cite{Seljak:1999zn}. In the absence of the cluster, the CMB is
smooth on arcmin scales, and so cluster lensing induces a dipole-like
signal aligned with the local background gradient of the temperature/polarization
anisotropies. Initially, subtraction of this background gradient to
measure directly the deflection field was
suggested~\cite{Vale:2004rh,Holder:2004rp}, but this proved
difficult. However, approaches based on the application of the
quadratic estimators designed for lensing by large-scale structure, or
on some modified version of them, have proved more
satisfactory on simulated data~\cite{Maturi:2004zj,Hu:2007bt,Yoo:2008bf}. Once the lensing
deflections have been reconstructed, the cluster mass can be extracted
optimally by application of a matched filter based on the expected
cluster profile (e.g., an NFW
profile~\cite{Navarro:1996gj})~\cite{Melin:2014uaa}. Alternatively,
cluster parameters can be estimated directly from the lensed CMB
fields with a parametric maximum-likelihood
approach~\cite{Lewis:2005fq}. Cluster mass estimation via CMB lensing
is particularly promising for large samples of high-redshift clusters,
where mass estimation by other means is very difficult.

Current high-resolution CMB observations are not sufficiently
sensitive to allow measurement of individual cluster masses via CMB
lensing. However, the mass scale of a cluster sample with a
sufficiently large number of elements can be estimated with moderate
$S/N$. Using data from SPT, the mass scale of 513 clusters was
estimated via a parametric maximum-likelihood
approach~\cite{Baxter:2014frs}, yielding results consistent with the
SZ-estimated mass scale and with the null hypothesis of no lensing
rejected at $3.2\,\sigma$. For \planck\ clusters, the approach proposed
in Ref.~\cite{Melin:2014uaa} was followed to estimate the hydrostatic
bias parameter $b$ that relates the X-ray derived mass $M_X$ and the
true mass $M_{500}$: $M_X = (1-b) M_{500}$~\cite{Ade:2015fva}. If the true mass is identified with a lensing-derived mass,
galaxy lensing prefers a low value for $1-b$, somewhere in between $0.6$
and $0.8$, which significantly relaxes the tension between the
observed cluster counts and those predicted in the $\Lambda$CDM model
with parameters determined from the primary CMB fluctuations.
However, CMB lensing
prefers a smaller bias (specifically, $1/(1-b) = 0.99 \pm 0.19$),
which goes in the opposite direction of increasing the tension with
the primary fluctuations.

The current applications of cluster lensing of the CMB only make use
of temperature observations. Indeed, for the noise levels of an
experiment like \planck, the $TT$ quadratic estimator has the lowest
reconstruction noise. However, for an experiment such as \mission,
the $EB$ estimator will be the most powerful, as in the case of
lensing by large-scale structure (Sec.~\ref{sec:recon}). This is significant
since polarization-based measurements should remove several
astrophysical sources of systematic error that complicate measurements
based on temperature. These include residual tSZ emission, the kinetic
SZ effect from cluster rotation (which induces a dipole-like signal with
a CMB frequency spectrum), and residual infrared emission from galaxies
within the cluster or along the line of sight.

\begin{figure}
\begin{center}
\includegraphics[width=12cm,angle=0]{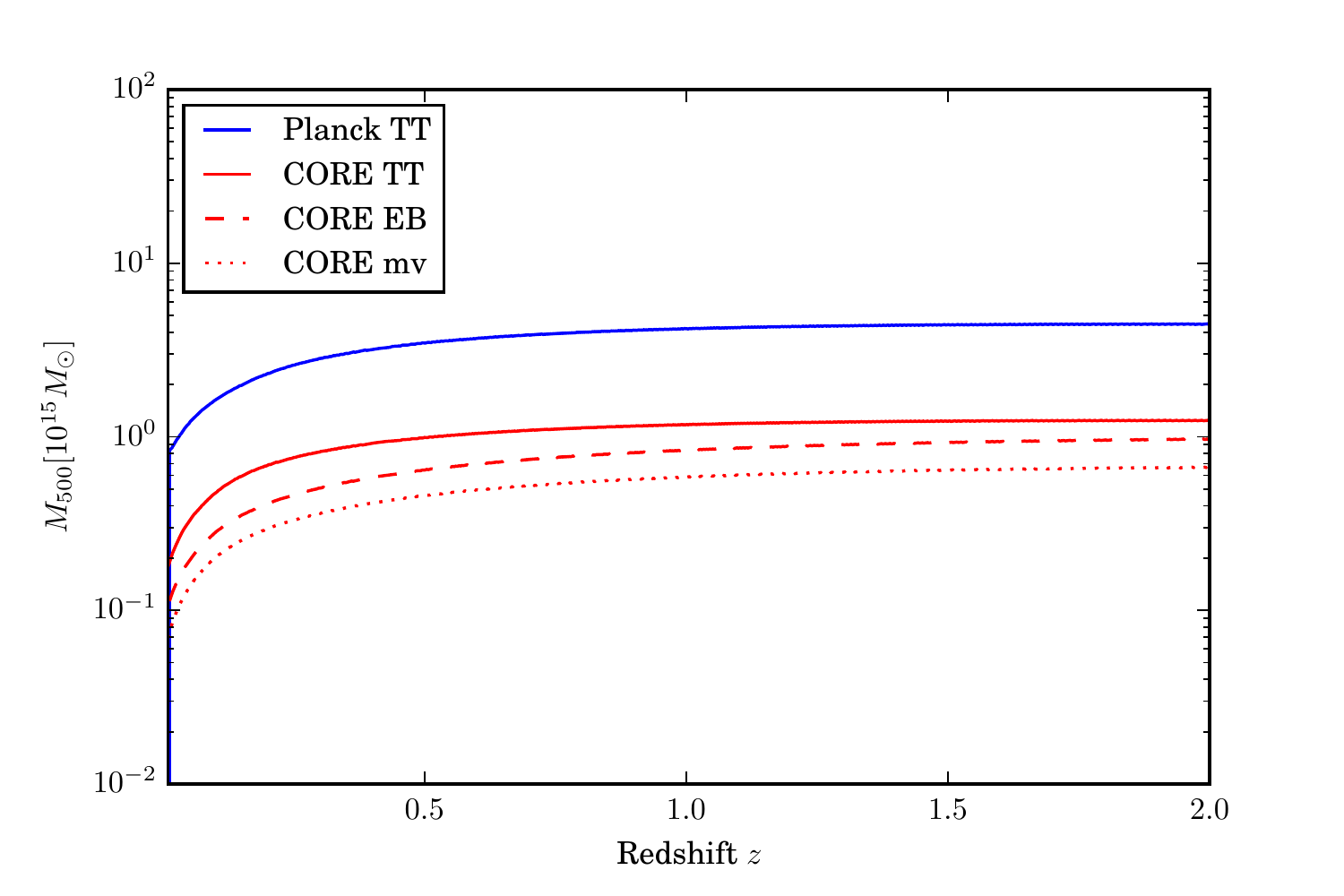}
\end{center}
\caption{
Limiting cluster mass as a function of redshift for which the $S/N$ on
a CMB lensing mass measurement with \mission\ is unity for an
individual cluster. Results are shown for the $TT$
(red solid), $EB$ (red dashed), and minimum-variance (red dotted)
quadratic estimators. For comparison, the equivalent for temperature
reconstructions at the sensitivity of \planck\ is also shown.}
\label{fig:clustermass}
\end{figure}

The potential of cluster mass measurements with \mission\ is
illustrated in Fig.~\ref{fig:clustermass}, which shows, as a function
of redshift, the minimum cluster mass with a mass measurement of $S/N
= 1$. Here, lensing is reconstructed with quadratic estimators using
multipoles $l \leq 3000$ in temperature and polarization, and the mass
estimated using the matched-filter method described
in Ref.~\cite{Melin:2014uaa}. The filter uses an NFW profile, truncated as
$5r_{500}$ (where $r_{500}$ is the radius at which the mass enclosed
is 500 times that for a uniform density equal to the critical density
at the cluster redshift).
Results are shown for the $TT$, $EB$, and
minimum-variance quadratic estimators. Similar results for $TT$ only
can be found in Ref.~\cite{Melin:2017lkr}.\footnote{%
The $TT$ results there are not directly comparable with those in
Fig.~\ref{fig:clustermass} due to several differences in
implementation, including the maximum multipole used in the
analysis.}
We use noise levels for the
combination of CMB channels used throughout this paper, assuming that
this is representative of the noise after astrophysical foreground removal.
We additionally propagate the effects of lensing by large-scale
structure (assumed independent of the cluster lensing) to the
forecasted errors on our mass measurements. For comparison with
previous experiments, results are also shown for a \planck-like
experiment with temperature noise levels of $45\,\mu{\rm K}\,{\rm
  arcmin}$ and a Gaussian beam with FWHM of 5
arcmin. Figure~\ref{fig:clustermass} shows that the $EB$ estimator is
more powerful than $TT$ for \mission, and that the improvement with
respect to \planck\ is significant. Individual cluster masses can be
measured with $S/N \geq 1$ for all clusters with $M_{500} > 10^{15}
\,M_\odot$ irrespective of redshift and over the full sky; the
accuracy is considerably better than this below $z=0.5$. While \mission\
lacks the resolution to be able to measure individual masses for
typical clusters that it will detect, the large sample size means that
scaling relations between tSZ observables and the true mass can be
accurately calibrated~\cite{Melin:2017lkr}. For example, assuming that the hydrostatic bias
parameter $b$ is independent of mass and redshift, \mission\ will be
able to calibrate this at the percent level using the clean $EB$ estimator.

\section{Impact of Galactic foregrounds on lensing reconstruction}
\label{sec:galforegrounds}

Polarized emission from Galactic dust is now known to be a major issue
for attempts to detect $B$-modes from primordial gravitational
waves~\cite{Planck2014-XXX,Ade:2015tva}. Any internal
reconstruction of the CMB lensing potential will also be contaminated
by residual foregrounds in the observed region of the sky. Since
lensing estimators rely on extracting the non-Gaussian signature in
the observed CMB that is characteristic of gravitational lensing (see
Sec.~\ref{sec:recon}), inherently non-Gaussian foreground fields such
as Galactic dust are of particular concern for lensing studies
\cite{Fantaye:2012}. Given that the $EB$ quadratic estimator will
provide most of the lensing information for \mission, characterising
the contamination of polarized dust emission to the recovered lensing
power spectrum is of vital importance. While most of the $S/N$ on lens
reconstructions at multipole $L$ from $TT$ and $EE$ come from squeezed
shapes, i.e., CMB modes at multipoles $l \gg L$, this is not true for
the more powerful $EB$ reconstruction except on the largest scales; a
significant fraction of the $S/N$ comes from $B$-modes with $l < 1000$
for any $L$~\cite{Pearson:2014qna}. As the dust $B$-mode power
spectrum is much redder than the $B$-mode spectrum from lensing, one
might expect dust to be a significant contaminant for $EB$
reconstructions at all multipoles $L$. Indeed, \planck\ 353\,GHz data show that over 70\,\% of the sky, the expected dust $B$-mode power at 150\,GHz exceeds the lensed $B$-mode power at all multipoles~\cite{Planck2014-XXX}.

For intermediate and high Galactic latitudes, the large-scale  ($40 <
l < 600$)  angular power spectra of the dust polarization are well
constrained by \planck\ observations at 353\,GHz, which are dominated
by polarized Galactic dust emission~\cite{Planck2014-XXX}. For
observations away from the Galactic plane, the dust power spectra at
353\,GHz are well-modelled by a single power law $C_l^{XX} =
A^{XX}_{\rm dust} (l/80)^\alpha$ for $X \in \{E,B\}$ and $\alpha =
-2.42 \pm 0.02$. Dust polarization arises from the alignment of
aspherical grains in the Galactic magnetic field (e.g.,
Refs.~\citep{Stein:1966, LeeDraine:1985, Planck2015XXI}). In
Ref.~\cite{Vansyngel:2016fbn}, Gaussian simulations of
the turbulent magnetic field in the Galaxy are used to argue that the power-law
slope $\alpha$ of the polarization angular power spectra directly
reflects the slope of the power spectrum of the turbulent
field. The polarized dust power-spectral amplitude $A^{EE}_{\rm dust}$
(given in $\mu {\rm K}^2_{\text{CMB}}$ at 353\,GHz) varies significantly
with sky coverage: $A^{EE}_{\rm dust} = 37.5\pm 1.6$ for the cleanest
24\,\% of the sky, while $A^{EE}_{\rm dust} = 328.0\pm2.8$ for the
cleanest 72\,\%, reflecting the large variation in dust column density
across the sky. The amplitudes $A^{BB}_{\rm dust}$ are approximately
half the corresponding $A^{EE}_{\rm dust}$ amplitudes.

A Gaussian and statistically-isotropic dust contribution would add
noise to the lensing reconstruction, but could be handled
straightforwardly and optimally using existing filtering
techniques. Note that such filtering requires knowledge only of the
total power spectra (including CMB, foregrounds and noise), which can
be approximated by smoothed versions of the measured spectra, and
fiducial lensed CMB power spectra. Such Gaussian and statistically-isotropic dust
would propagate no bias into the lensing power spectrum
or derived parameters. The statistical anisotropy of dust can
also be handled with existing techniques if reconstructions are made
locally. Using a realisation-dependent calculation of the Gaussian
noise-bias $N_L^{(0)}$ would mitigate against small errors in simulating the dust locally as being statistically
isotropic, with power spectra calibrated within each patch~\cite{Namikawa:2013}.
Rather, it is the non-Gaussianity of dust emission that is
particularly problematic for lensing studies. The alignment of dust
grains, which sources the polarized emission, and their spatial
distribution are highly complex and imperfectly modelled. There is
large variation in the Galactic magnetic field orientation along any
given line of sight, and this leads to scatter in the polarization
fraction~\cite{Ghosh:2016}. The trispectrum of the dust emission will
bias estimates of the lensing power spectrum, with the bias going
like the fourth power of the polarized dust amplitude. 

One hindrance to quantifying the dust bias to lensing is that the
small-scale polarization field of the dust is not well constrained by
current data. For future surveys, the extent of the dust contamination
of lensing estimators is therefore not well known; there is
considerable model uncertainty about the expected amplitude and shape
of the dust four-point signal. Here, we consider dust maps constructed
for the \planck\ FFP8 simulations~\cite{Planck2015XII}. While these
are well motivated in that they are data-derived (principally from
\planck\ 353\,GHz observations), the polarization observations in
particular are extremely noisy at small scales.\footnote{The
  construction of the simulated maps is described in detail in
  Ref.~\cite{Planck2015XII}. The degree and orientation of dust polarization
  on scales larger than $0.5\,\text{deg}$ are derived from the ratio
  of smoothed maps of the Stokes parameters $Q$ and $U$ at 353\,GHz
  and the GNILC-reconstructed map of the dust total intensity. These
  ratio maps are extended to smaller scales with a Gaussian
  realisation. Finally, the full-resolution polarization maps are
  obtained by multiplying with the total intensity GNILC dust
  template. The small-scale dust polarization inherits non-Gaussianity
  from the total intensity, but does not properly reflect the
  non-Gaussian structure due to the small-scale Galactic magnetic field.}
These tracer maps
therefore cannot be fully representative of the small-scale dust
polarization on the sky that will be seen by \mission. Another approach
under investigation is to derive a dust polarization tracer map from
observations of Galactic neutral hydrogen ({\sc Hi}), for which the
filamentary structure has been shown to correlate strongly with the
orientation of the Galactic magnetic field, and hence the dust
polarization angle~\cite{Clark:2015, Ghosh:2016}. Whether these
different dust tracers yield similar inferences about the lensing dust
bias is the subject of ongoing work. As a space-based CMB experiment,
\mission\ will allow for precision observations at high frequencies --
something that is unattainable from the ground, and therefore an
obvious concern for ground-based experiments such as CMB-S4. \mission\
observations would thereby eliminate the data deficiency on the
small-scale dust polarization, allowing for a direct measurement of
dust bias and, more practically, providing the high-frequency data
needed for dust subtraction through component-separation techniques
(see Ref.~\cite{Remazeilles:2017szm} for a detailed analysis of component separation
for \mission).

Given that the dust power spectra fall rapidly with multipole, one way
to reduce dust contamination is to exclude large-scale modes from
the lensing reconstruction analysis~\cite{Fantaye:2012}. In the choice
of the minimum multipole $l_{\rm min}$ to use in the analysis, there
is a trade-off between bias and variance: including more modes will
reduce the sample variance, but the filtered maps will be
correspondingly contaminated by non-Gaussian large-scale dust modes
that will propagate into a lensing bias. As discussed above, the
strong spatial variation of the dust signal may suggest performing
reconstructions locally. The choice of the minimum multipole $l_{\rm min}$ to use in the analysis may vary depending on the position of observed patch relative to the Galaxy. Furthermore, the lensing signal is extracted through weighted combinations of filtered versions of the observed fields, with the optimal filters themselves dependent on the local dust power spectrum, again suggesting a local approach to lensing reconstruction.

As well as signals from our Galaxy, non-Gaussian extragalactic
foregrounds can bias lens reconstruction. Potential biases for the
$TT$ quadratic estimator are studied in
Ref.~\cite{vanEngelen:2013rla}. For temperature observations, bright
galaxies (flux density $F_{150\,{\rm GHz}} \gtrsim 1$\,mJy) and
massive galaxy
clusters ($M\gtrsim10^{14}M_\odot$) must be appropriately masked, and
a maximum multipole $l_{\rm max} \approx 2500$ used for lensing
reconstruction, to reduce the induced bias to acceptable levels for
\mission. The extra-Galactic contamination in polarization is not
robustly quantified, but is expected to be less problematic due to the
typically low polarization fraction of these sources~\cite{Smith:2008an}.

\subsection{Quantifying lensing bias from Galactic dust}

\begin{figure}[p]
\centering
\includegraphics[width = 0.49\textwidth]{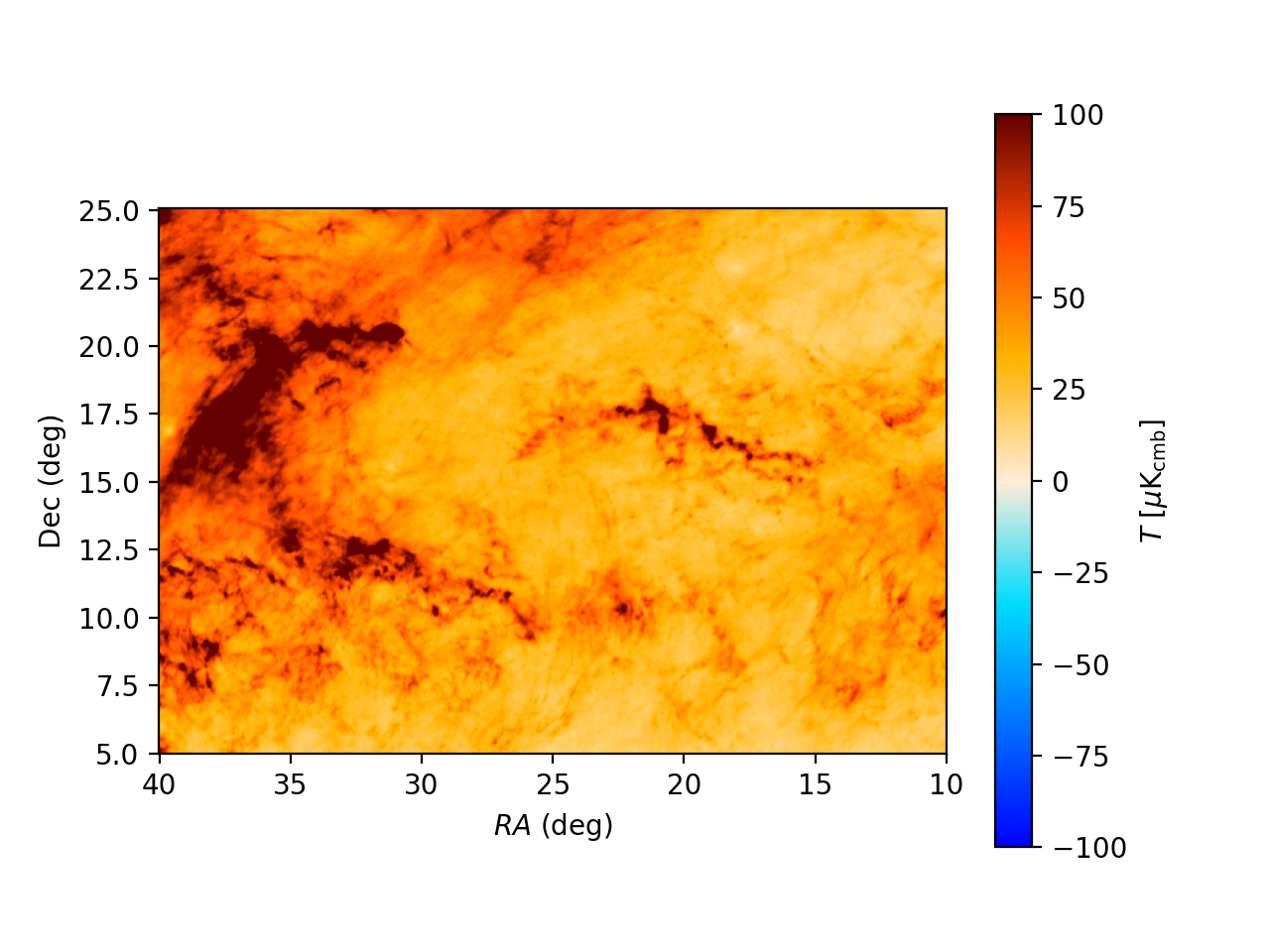}
\includegraphics[width = 0.49\textwidth]{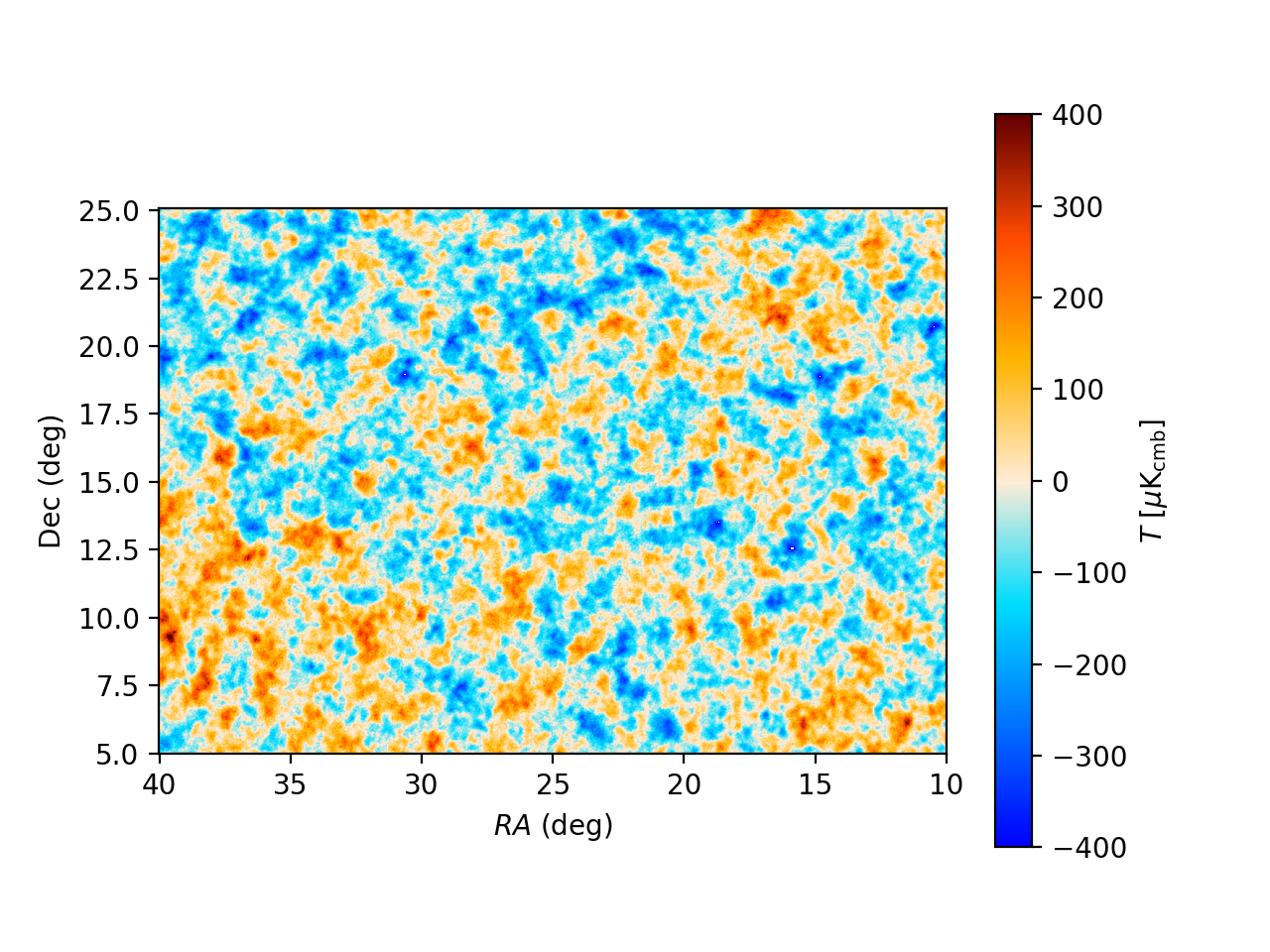}
\includegraphics[width = 0.49\textwidth]{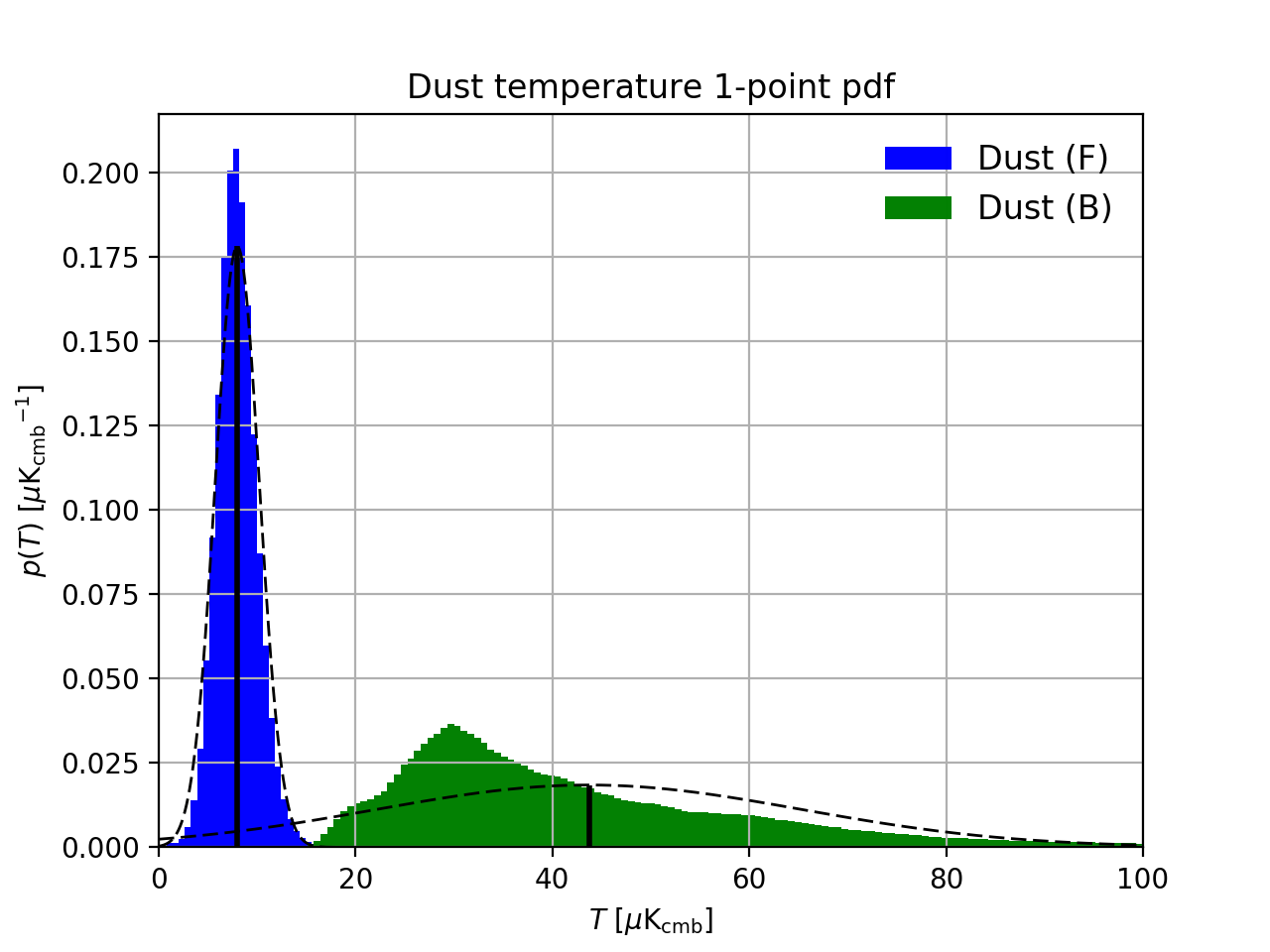}
\includegraphics[width = 0.49\textwidth]{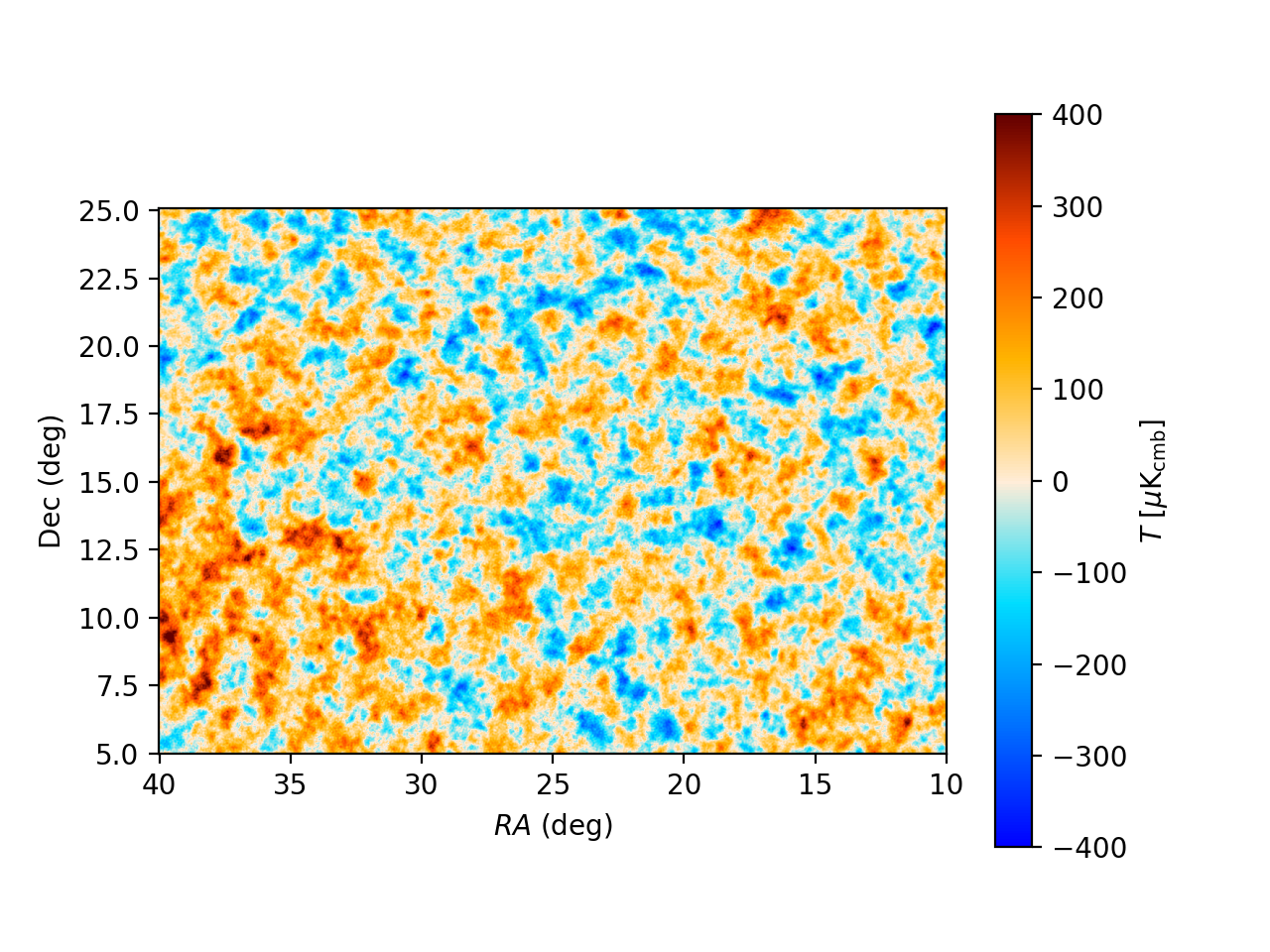}
\centering
\includegraphics[width = 0.49\textwidth]{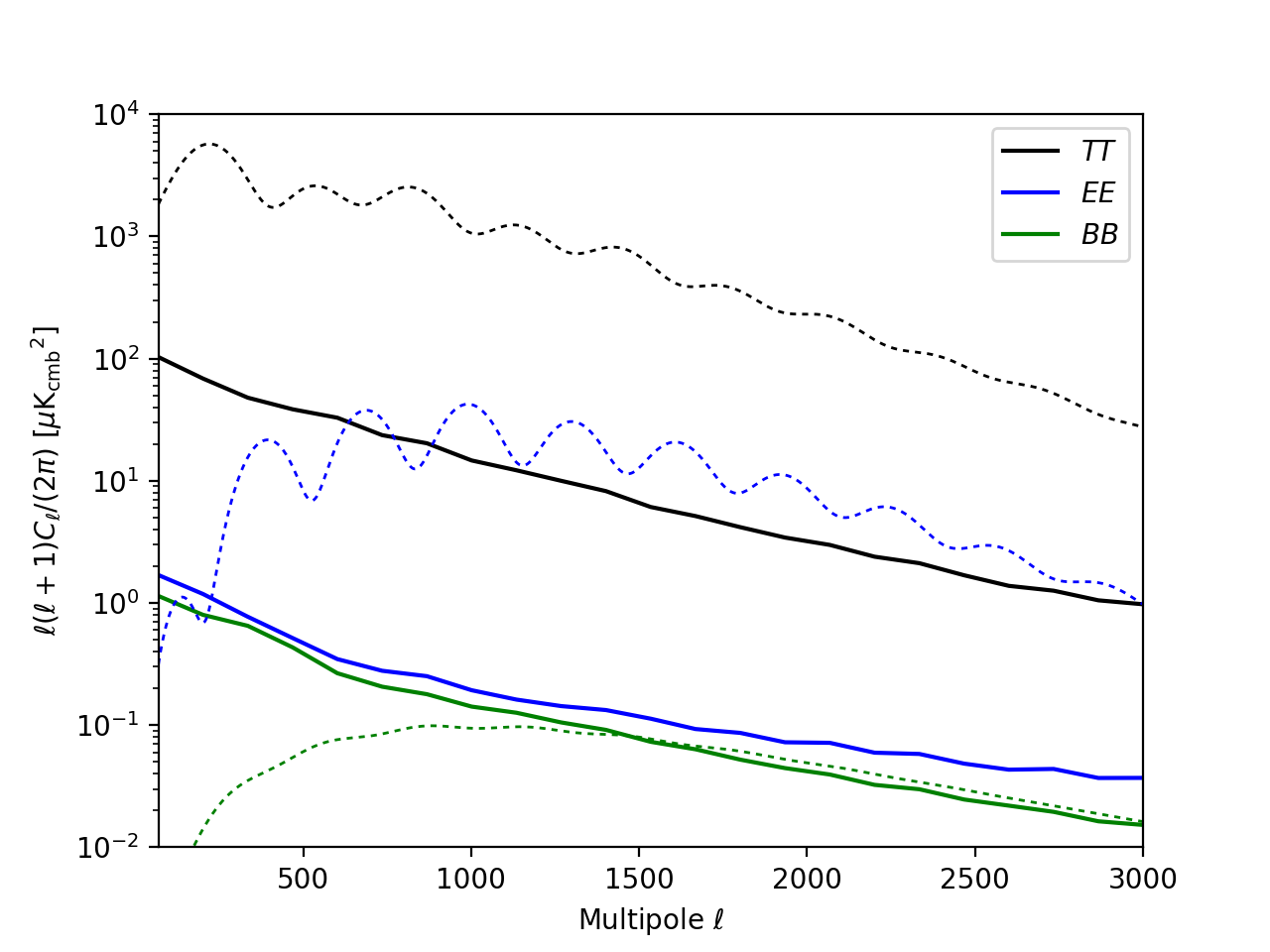}
\caption{{\it Top left}: Dust emission in total intensity at 150\,GHz in the
  analysis field. The diffuse, non-Gaussian nature of the dust
  emission is clear. For comparison, the r.m.s. CMB fluctuations are
  around $70\,\mu\text{K}$. {\it Middle left}: Pixel histogram
  (1-point function) for the dust intensity. The field used in this
  analysis is labelled `B' and coloured green; the non-Gaussianity of
  the 1-point density is apparent when compared against the
  black-dashed Gaussian distribution that has the same mean and
  dispersion. The field labelled `F' and coloured blue comes from the
  cleanest $600\,\text{deg}^2$ patch of dust emission near the
  Southern Galactic pole. The CMB temperature (\planck\ {\tt SMICA}
  map) in the analysis region is also shown without ({\it top right}) and with ({\it
    middle right}) the additive dust emission component; the effect of
  dust at 150\,GHz is visible by eye in this region. {\it Bottom}:
  Dust power spectra in the analysis patch (solid lines). CMB
  fluctuations (dashed lines) dominate the variance in temperature,
  but in polarization the $BB$ dust power spectrum is greater than or
  comparable to the
  CMB spectrum for all scales, while the $EE$ dust spectrum exceeds
  the CMB spectrum on large scales.}
\label{figDustT}
\end{figure}

The broad frequency coverage of \mission\ will allow for accurate
foreground subtraction through component separation
techniques~\cite{Remazeilles:2017szm} and a corresponding mitigation of the lensing bias.
Here, we attempt to bound the overall bias to the lensing potential
power spectrum by performing lensing reconstruction on CMB simulations
over a field of 600\,$\text{deg}^2$, to which a fixed realisation of bright dust emission
is added in temperature and polarization, with no component separation
performed. The field chosen has the brightest dust emission of all
regions of this size that lie outside the \planck\ analysis mask, and
so likely will be included in the \mission\ lensing analysis; the
results therefore represent a rough upper bound on the dust
contamination. The dust tracer map is more reliable in regions of
bright emission because of the lower fractional contamination from
residual CIB or detector noise. The dust intensity at 150\,GHz in
this field is shown in Fig.~\ref{figDustT} along with its
corresponding 1-point distribution. For comparison, we also show the
1-point function of dust intensity emission of the cleanest 600\,$\text{deg}^2$ region, near the Southern Galactic pole, in which the root-mean-square intensity fluctuations are an order-of-magnitude below our analysis field.

\begin{figure}[t]
\includegraphics[width = 0.49\textwidth]{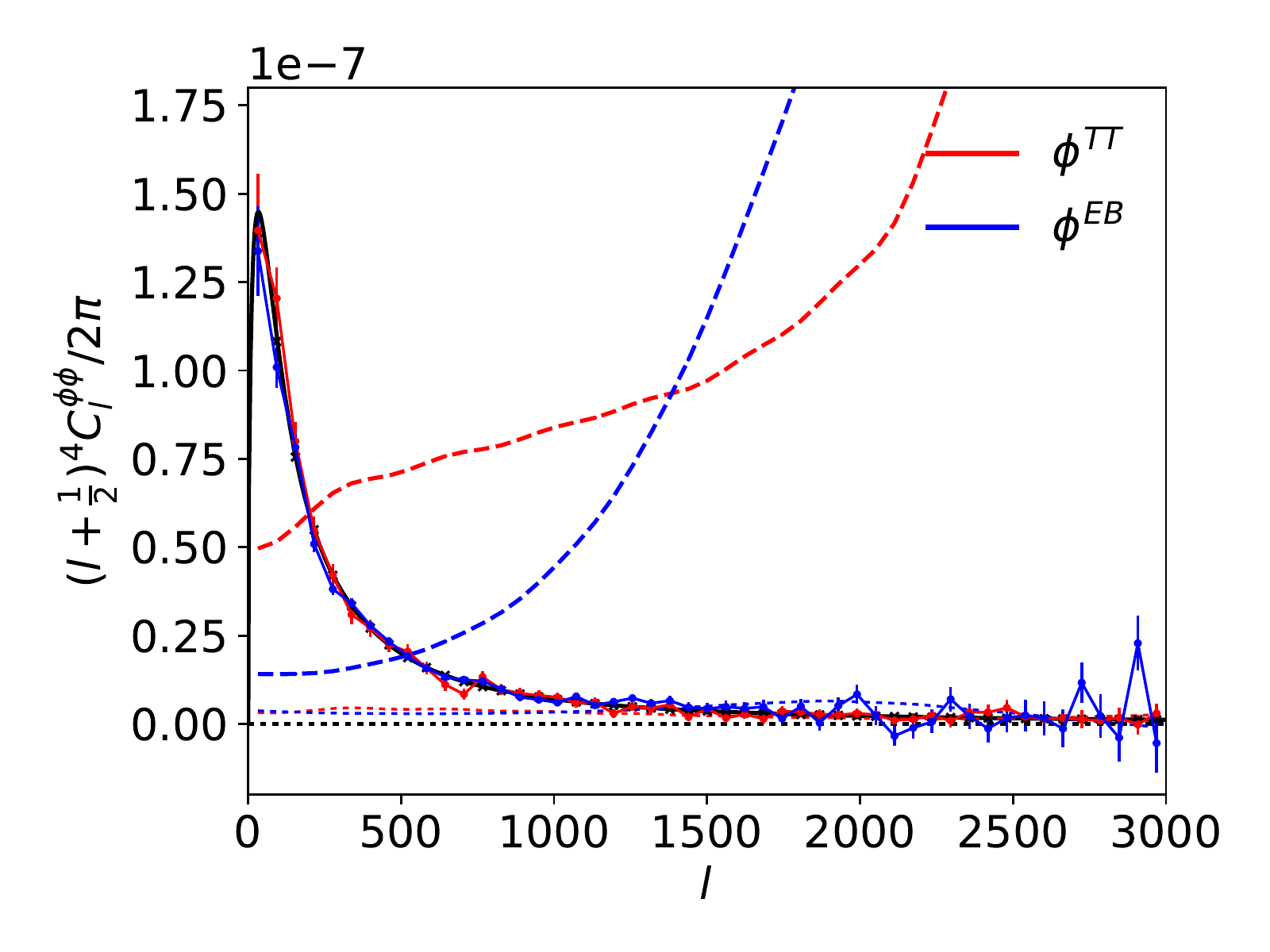}
\includegraphics[width = 0.49\textwidth]{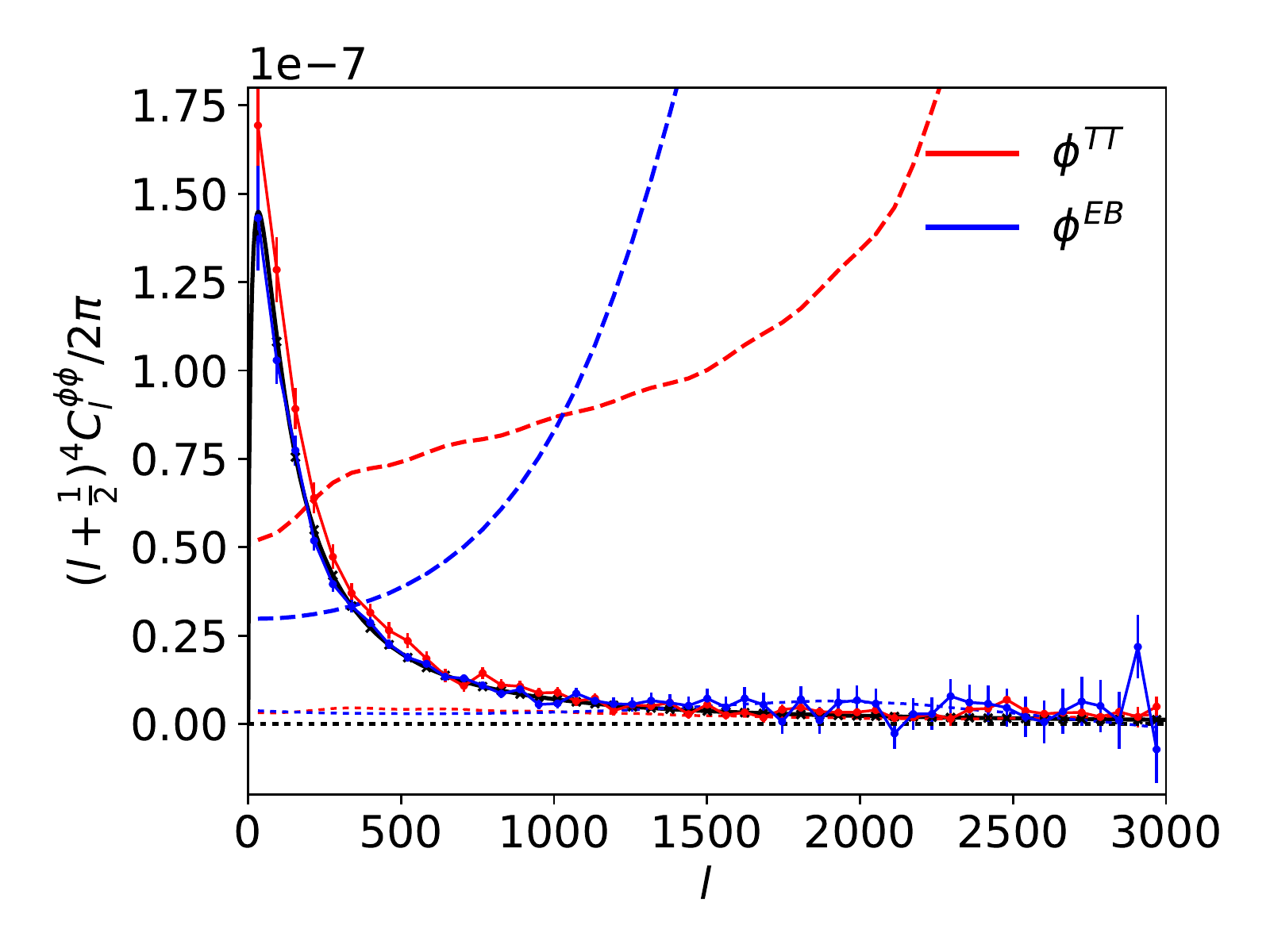}
\includegraphics[width = 0.49\textwidth]{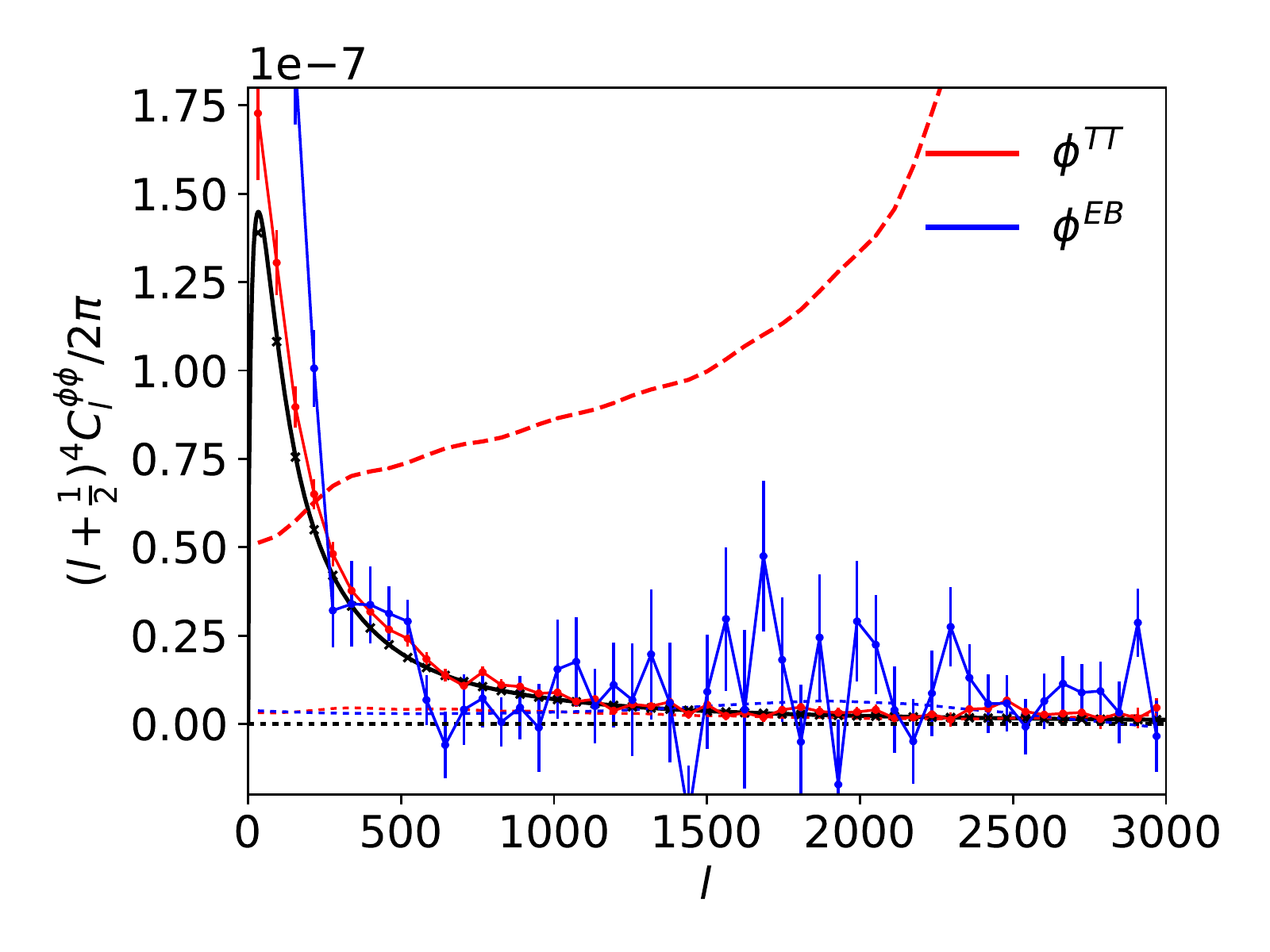}
\centering
\caption{{\it Top left}: Mean lensing power spectrum reconstructions
  over five mock \mission\ observations at 150\,GHz, including
  realistic noise, with no dust contamination. The lensing power
  spectrum (black solid line)  is recovered in an unbiased fashion
  (see also Table~\ref{tabLensingAmp4}). Dashed lines show the $N^{(0)}_L$
  Gaussian noise power in the reconstructions, and dotted lines are
  the small $N_L^{(1)}$ biases. The measured spectra are corrected for both
  such biases. The lensing detection significance of the $EB$
  reconstruction has about twice the power of $TT$ in the no-dust
  case. {\it Top right}: As top left, but with the dust field added
  before reconstruction, and including the dust power in the lensing
  filters. The bias from dust is clear by eye in the temperature
  reconstruction (red points). With this filtering for \mission, we
  expect a roughly 15\,\% bias for the $TT$ estimator and negligible
  bias for $EB$. Note that these results are specific to this bright
  dust field, and assume no Galactic foreground removal. It can be seen that uncertainties are inflated relative to the no-dust case due to the additional variance of the dust component.
{\it Bottom}:
  As top right, but without including the dust power in the
  filtering. In this case the $EB$ reconstruction is highly
  sub-optimal (the $N^{(0)}_L$ Gaussian noise power is off the scale
  of the plot) and biased. There is little change in the $TT$
  reconstruction as the dust power is subdominant to the CMB power. In
  all panels, the error bars are analytic estimates from $N_L^{(0)}$
  and $C_L^{\phi\phi}$ and have been scaled to reflect $f_{\rm sky}=0.7$.
  The same CMB and noise realisations are used in all panels.}
\label{figCORENoise}
\end{figure}

The bias to the lensing reconstruction from dust emission is
quantified as follows. Appropriately-correlated Gaussian CMB
realisations on the flat sky are drawn from fiducial spectra based on
the \planck\ best-fit cosmological
parameters~\cite{Planck2014-XXX}. The temperature and polarization
fields are lensed by remapping with the gradient of a Gaussian lensing potential, itself drawn from the
fiducial lensing potential power spectrum. The fixed dust realisation
and scale-dependent Gaussian noise appropriate for the \mission\ CMB channels\footnote{%
 The
  noise level is determined from the combination of \mission's CMB
  channels, i.e., the six channels in the range 130--220\,GHz,
 used for forecasts throughout this paper. The assumed dust
  level in the field has not been corrected for (weighted) averaging the dust spectral energy
  distribution across these channels; rather it is simply the emission
  at 150\,GHz.} are added to these lensed simulations. 
The lensing potential is then reconstructed using flat-sky
implementations of the quadratic estimators~\cite{Hu:2001kj} from the
{\tt quicklens} package.\footnote{\url{https://github.com/dhanson/quicklens}}
When filtering the CMB fields, $X \rightarrow \bar{X}$ [see
Eq.~\eqref{eq:quadest}], during lens reconstruction, we include 
a model dust power spectrum, obtained as a power-law fit to the dust
power spectra in this particular region on the sky. Failure to include
the dust power in the filtering can lead to biased and sub-optimal
reconstructions (see Table~\ref{tabLensingAmp4}). The minimum CMB
multipole used is $l_{\rm min} = 12$ corresponding to the longest
non-constant mode supported on the patch.
The mean auto-power spectra of the $TT$ and $EB$ reconstructions are
shown in Fig.~\ref{figCORENoise}. The realisation-dependent
$N^{(0)}_L$ bias and an analytic approximation of the sub-dominant
$N_L^{(1)}$ bias have been subtracted from the raw power spectrum to
obtain an unbiased estimator (in the absence of non-Gaussian
foregrounds) of the underlying lensing power. 

We note that here we scale the 353\,GHz dust emission to an effective
150\,GHz observing channel using a modified black body spectrum with
temperature $T_{\rm dust} = 21\,{\rm K}$ and spectral index
$\beta_{\rm dust} = 1.5$. We add this scaled dust component to
noisy, lensed CMB maps with noise level appropriate to the combination of \mission's CMB
  channels, i.e., the six channels in the range 130--220\,GHz,
 used for forecasts throughout this paper. This procedure is more akin
 to how a ground-based experiment with sensitivity around $2\,\mu
 \text{K}\,\text{arcmin}$ at 150\,GHz would observe and analyse the
 CMB sky. Since dust emission rises strongly with frequency, the
 actual level of dust emission in the six-channel combination, based
 on inverse-noise-variance weighting, would be around a factor of
 $1.5$ higher than at 150\,GHz assuming no component separation. 

We distill the effect of the dust bias by fitting a lensing amplitude
parameter $A$, which scales the amplitude of the fiducial lensing
power spectrum, to the estimated power spectrum.
Any statistically-significant deviation from
unity in this parameter ($A \ne 1$) represents biasing from the dust
emission which, if unmodelled, would directly impact any cosmological
inference from the lensing measurement. For example, the effect of
neutrino mass is to suppress the lensing power spectrum
(Sec.~\ref{sec:LSS}); a bias in the lensing amplitude $A$ would therefore directly propagate into crucial
cosmological parameters. To keep the bias well below the statistical
error on a measurement of the lensing amplitude, we require biases
below $O(0.1)\,\%$ for an $EB$-based analysis.

\begin{table}[t]
\begin{center}
\begin{tabular}{c c c c}
\hline
\hline
                        & No dust   & Bright dust field   & Bright dust field \\
                        &           &               & (inc. dust power
                                                      in filters)  \\
\hline
         $TT\times TT$  & $A = 1.002 \pm 0.008$ & $A = 1.169 \pm 0.008$ & $A = 1.158 \pm 0.008 $ \\
         $EB\times EB$  & $A = 0.997 \pm 0.004$ &$A = 1.615 \pm 0.030$
                                                          & $A = 0.999
                                                            \pm 0.006
                                                            $ \\
\hline
            \end{tabular}
\end{center}
\caption{Fits to the lensing power spectrum amplitude $A$ for mock
  \mission\ observations at 150\,GHz (see
  Fig.~\ref{figCORENoise}). Fits are performed over the multipole
  range $2 \leq L < 3000$ for the recovered lensing power
  spectrum from the $TT$ and $EB$ estimator. The central value quoted
  is the mean over five simulations,
  while the error is appropriate to a single realisation and has been
  scaled to reflect a full-sky analysis ($f_{\rm sky} = 0.7$). The
  standard error on the mean of $A$ is a factor $3.1$ larger than the
  errors shown. We see
  that without dust the input amplitude can be recovered to high
  accuracy. Performing the reconstruction over a dusty region induces
  bias in both estimators and additional variance for $EB$. When
  including the dust power in the lensing filters, the bias in the
  power from the $EB$ estimator is removed, but a 15\,\% bias remains
  for the $TT$ estimator since including the dust power makes only a
  small change to the temperature filter on the (small) scales that dominate
  the $TT$ reconstruction. Note that no foreground removal is assumed
  in this analysis.}
\label{tabLensingAmp4}
\end{table}

The principal results of this analysis are shown in
Table~\ref{tabLensingAmp4}. For this region of the sky, with its
atypically bright dust emission, the bias to \textit{CORE}
observations is around 15\,\% in the temperature reconstruction, but
much smaller for $EB$ (below the approximately 2\,\% level to which we
have sensitivity with only the five simulations used here).
This clarifies the need for
\textit{CORE} to perform lensing reconstruction on
foreground-subtracted temperature maps. The addition of dust has
little effect on the lensing detection significance for the $TT$
estimator since the dust power is small compared to the CMB
power. For the same reason, including dust power in the lensing
filters is ineffective in mitigating the dust bias for temperature
reconstructions. In this case, explicit high-pass filtering of the
data may be more effective. In contrast, the dust power is
comparable to, or larger than, the CMB polarization power across a
wide range of scales, and therefore has a stronger effect on the lensing
reconstruction uncertainties. These numerical conclusions were derived
assuming that the model dust maps used are representative of the true
dust emission on the sky; the present paucity of data on
high-resolution dust polarization limits the scope to remove this caveat in the near future.

We also investigate how the lensing bias from dust is mitigated by
reducing the amplitude of the dust emission by hand in the
simulations. This process can be regarded as modelling the dust
residuals after foreground cleaning, with the residual amplitude in
the cleaned map expressed as a fraction of the dust emission at
150\,GHz. We consider various levels of residual dust contamination
(Table~\ref{tabResiduals}), and here do not include any dust power in
the lensing filters, reflecting a global analysis. We see that, even in
this particularly bright field, the
dust bias can be reduced to acceptable levels for \mission\ (in both
temperature and polarization reconstructions) if foreground cleaning
can reduce the residual dust contamination to around 10\,\% of the
amplitude of the raw
emission at 150\,GHz (1\,\% in power).

\begin{table}[t]
\footnotesize{
\begin{center}
\begin{tabular}{c c c c c c c c c c }
\hline
\hline
                        & No dust   & Bright dust   & 50\,\% dust & 20\,\% dust & 10\,\% dust & 1\,\% dust\\
\hline
         $TT\times TT$  & $1.002 \pm 0.008$ & $1.169 \pm 0.008$ & $1.025 \pm 0.008$ & $1.007 \pm 0.008 $ & $1.004 \pm 0.008 $ & $1.002 \pm 0.008 $\\
         $EB\times EB$ & $0.997 \pm 0.004$ & $1.615 \pm 0.030$ & $1.004 \pm 0.010$ & $0.982 \pm 0.005 $ & $0.994 \pm 0.004 $ & $0.998 \pm 0.004 $ \\
\hline
            \end{tabular}
\end{center}
}
\caption{Fits to the lensing amplitude $A$ for mock \mission\
  observations over the bright dust field, as
  in Table~\ref{tabLensingAmp4}, with the amplitude of the (residual) dust emission reduced
  down to the indicated percentage of the raw emission at
  150\,GHz. The dust power is not included in the lensing filters
  here, reflecting a global analysis.}
\label{tabResiduals}
\end{table}

\mission\ will characterise the Galactic dust accurately -- both in
its spatial variation and in its spectral energy distribution --
through its high-sensitivity, multi-frequency coverage and its high
resolution in the dust-dominated channels. With component separation
techniques this will allow for the construction of
foreground-subtracted maps~\cite{Remazeilles:2017szm} on which lensing reconstruction can be
performed, mitigating the dust bias. Ongoing work aims to quantify the
residual lensing bias after foreground cleaning, accounting for the
\mission\ specification of frequency channels and map-level sensitivities.

\section{Conclusions}
\label{sec:conclusions}

Weak gravitational lensing of the CMB has
great potential as a relatively clean probe of the large-scale clustering of
all mass to high redshift. \mission\ has been designed so that it is
able to exploit much of
this potential. We discussed how the lensing map reconstructed from
\mission\ data would have statistical power significantly extending
what can be achieved with the current generation of
experiments. Lensing impacts many of the science goals of \mission;
here we have highlighted its role in measuring the absolute neutrino
mass scale, the growth of structure across cosmic time through
cross-correlation with other large-scale structure surveys, delensing
$B$-modes in searches for the polarization signal of primordial gravitational
waves, and calibration of cluster masses for accurate interpretation
of counts of galaxy clusters across redshift.

Current CMB lensing reconstructions are dominated by the temperature
anisotropies. However, the lensing information that can be extracted from the
temperature is severely limited by its Gaussian fluctuations, which
add an irreducible noise to the reconstruction. Since $B$-mode polarization on
intermediate and small scales is only expected to be produced by
gravitational lensing, polarization-based reconstructions can
circumvent this limitation. Moreover, while
the interpretation of temperature-based reconstructions needs to take careful account of
non-Gaussian extragalactic foregrounds, polarization-based
reconstructions are expected to be much cleaner.
However, achieving a precise lens reconstruction with polarization requires sufficient
sensitivity and resolution to image the faint lens-induced $B$-modes over a
broad range of scales. One of the main science goals of \mission\ --
searching for the $B$-mode polarization from primordial gravitational
waves down to tensor-to-scalar ratios $r \sim O(10^{-3})$~\cite{Finelli:2016cyd} -- already
demands noise levels below the lens-induced $B$-modes on degree
scales. By combining this sensitivity with angular resolution of around
$6\,\text{arcmin}$, \mission\ is able to reconstruct lensing via the
$EB$ estimator over the full sky with $S/N$ greater than unity
\emph{per mode} for lens multipoles $L < 500$. \mission\ is therefore uniquely able, amongst currently-proposed satellites,
to delens its measured degree-scale $B$ modes with an internal lens
reconstruction. Generally, delensing requires a high-$S/N$ proxy for the CMB
lensing potential and high-$S/N$ $E$-mode measurements both over a broad
range of scales. We showed that \mission\ would be able to reduce the
power of lens-induced $B$-modes by around 60\,\% with internal
delensing. A similar level of delensing should be possible with the
clean measurement of the cosmic infrared background from the
multi-frequency \mission\ data, providing a valuable cross-check on
results with internal delensing. CIB delensing is particularly helpful
for small-scale lenses where the statistical noise on lens
reconstructions becomes large. Indeed, the optimal combination of an
internal lens reconstruction and the CIB can reduce the lensing
$B$-mode power to around 70\,\%. In the null hypothesis, $r=0$, this
would improve the error on the tensor-to-scalar ratio by a factor
$2.5$ compared to no delensing. 

Similar lensing performance to \mission\ could also be achieved with a
future ground-based survey, e.g.,
CMB-S4~\cite{Abazajian:2016yjj}.
Improved angular resolution relaxes the noise requirement a little, but for
similar statistical power one still needs
polarization sensitivity better than $3\,\mu\text{K}\,\text{arcmin}$
(at 1\,arcmin resolution, for example) over nearly the full sky. To reach this
sensitivity below the atmosphere requires roughly two orders of
magnitude more detectors than on \mission. For the goal of measuring
neutrino masses with CMB lensing, a critical limitation arises from
uncertainty in the optical depth to reionization, $\tau$. It is
currently unknown, however, whether it will be possible to measure this
parameter precisely with large-angle $E$-mode measurements from
sub-orbital experiments.
We infer the neutrino
mass by its impact on the growth of structure from high redshift, as
measured with the primary CMB fluctuations, to lower redshifts, as
measured by CMB lensing. Our knowledge of the amplitude of the
primordial fluctuations, $A_{\rm s}$, from the primary CMB is limited by uncertainty
in the optical depth since only the combination $A_{\rm s} e^{-2\tau}$
is well determined. With lensing measurements of the precision
expected from \mission, the total neutrino mass can be measured to a
precision of $17\,{\rm meV}$, when combined with contemporaneous BAO
distance measurements, \emph{provided that the optical depth is also
  measured to cosmic-variance limits}. This compares to the minimal
mass implied by neutrino oscillations of approximately $60\,{\rm
  meV}$. \mission\ is designed so that it can make precision
measurements of the reionization feature in large-angle polarization
and so determine $\tau$ to the cosmic-variance limit $\sigma(\tau)
\approx 0.002$. In contrast, if we had to rely on the current \planck\
determination, with $\sigma(\tau) \approx
0.009$~\cite{Aghanim:2016yuo}, the error on the total neutrino mass
would almost double to $30\,{\rm meV}$ and a detection would not be guaranteed.

Finally, we note that \mission\ is designed with broad frequency
coverage so that it can accurately separate the CMB from Galactic and
most extragalactic foreground emission~\cite{Remazeilles:2017szm}. We know from current attempts
to measure degree-scale $B$-modes that accurate removal of Galactic
dust is critical, even in the cleanest parts of the
sky~\cite{Ade:2015tva}. Lensing reconstruction mostly relies on smaller-scale
modes of the CMB so the expectation is that foreground cleaning will
be less demanding than for degree-scale $B$ modes. We have attempted
to quantify this, presenting some preliminary results on the level of bias that
would arise from (non-Gaussian) residual Galactic dust contamination in lensing power
spectrum measurements. Even in the regions of brightest emission away
from the Galactic plane, cleaning that suppresses the
dust emission amplitude to 10\,\% of the raw emission at 150\,GHz is sufficient
to reduce the bias to acceptable levels. Cleaning to such levels
should be achievable with \mission, and, of course, the demands are
less stringent in regions with more typical levels of
emission. 

\appendix
\section{Quadratic lensing reconstruction}
\label{app:quadrecon}

The linear response of the covariance between lensed CMB fields
$\tilde{X}_{l m}$ and $\tilde{Y}_{lm}$, where $X$ and $Y = T$, $E$, or $B$, to a variation
in the lensing potential is
\begin{equation}
\langle \delta (\tilde{X}_{l_1 m_1} \tilde{Y}_{l_2 m_2})
\rangle \approx \sum_{LM} (-1)^M \left(
\begin{array}{ccc}
l_1 & l_2 & l_3 \\
m_1 & m_2 & -M
\end{array}
\right) \mathcal{W}^{XY}_{l_1 l_2 L} \delta \phi_{LM} \, .
\end{equation}
The response functions are 
\begin{align}
\mathcal{W}^{TT}_{l_1 l_2 L} &= C_{l_2}^{TT} {}_+ F^0_{l_1 L l_2} +
                               C_{l_1}^{TT} {}_+ F^0_{l_2 L l_1} \, ,
  \\
 \mathcal{W}^{EE}_{l_1 l_2 L} &= C_{l_2}^{EE} {}_+ F^2_{l_1 L l_2} +
                               C_{l_1}^{EE} {}_+ F^2_{l_2 L l_1} \, ,
  \\
\mathcal{W}^{TE}_{l_1 l_2 L} &= C_{l_2}^{TE} {}_+ F^0_{l_1 L l_2} +
                               C_{l_1}^{TE} {}_+ F^2_{l_2 L l_1} \, ,
  \\
\mathcal{W}^{TB}_{l_1 l_2 L} &= i C_{l_1}^{TE} {}_- F^2_{l_2 L l_1} \, ,
  \\
\mathcal{W}^{EB}_{l_1 l_2 L} &= i C_{l_1}^{EE} {}_- F^2_{l_2 L l_1} \, ,
\end{align}
where $C_l^{XY}$ are the lensed spectra, and we have defined
\begin{multline}
{}_\pm F^s_{l_1 L l_2} = \frac{1}{2}\left(1 \pm (-1)^{l_1 + l_2 +
    L}\right) \left[L(L+1)-l_1 (l_1+1) + l_2 (l_2+1)\right]
   \\ \times 
\sqrt{\frac{(2L+1)(2l_1+1)(2l_2+1)}{16\pi}} \left( \begin{array}{ccc}
l_1 & L & l_2 \\
s & 0 & -s \end{array} \right)\, .
\end{multline}
Note that the ${}_+ F$ vanish for $l_1+l_2+L$ odd, and the ${}_- F$ for
$l_1+l_2+L$ even. The response function $\mathcal{W}^{XY}_{l_1 l_2 L}$
are non-zero only for $l_1+l_2+L$ even for parity-even combinations,
such as $TT$ or $TE$, while they are non-zero only for $l_1+l_2+L$ odd
for parity-odd combinations, such as $TB$ and $EB$. Moreover, they are
real for parity-even combinations and imaginary for odd parity, and
satisfy $\mathcal{W}^{XY}_{l_1 l_2 L} = (-1)^{l_1+l_2+L}
\mathcal{W}^{YX}_{l_2 l_1 L}$.

The optimal quadratic estimator was given in Eq.~\eqref{eq:quadest}, which
we repeat here for convenience:
\begin{equation}
\hat{\phi}^{XY}_{LM} = \frac{(-1)^M}{2}\frac{1}{\mathcal{R}_L^{XY}} \sum_{l_1 m_1,
  l_2 m_2} \left(
\begin{array}{ccc}
l_1 & l_2 & L \\
m_1 & m_2 & -M
\end{array}
\right)[\mathcal{W}^{XY}_{l_1 l_2 L}]^\ast \bar{X}_{l_1 m_1} \bar{Y}_{l_2
  m_2} \, .
\end{equation}
Throughout this paper, we assume that the temperature and polarization
fields are filtered independently, following
Ref.~\cite{Ade:2015zua}. In this case, for an isotropic survey the
inverse-variance-filtered fields $\bar{X}_{lm} = F_l^X X_{lm}$, where
the filter is the inverse of the total power spectrum: $F_l^X =
1/C_{l,\text{tot}}^{XX}$. In this case, the normalisations of the
quadratic estimators are
\begin{equation}
\mathcal{R}_L^{XY} = \frac{1}{2(2L+1)}\sum_{l_1 l_2} F_{l_1}^X
F_{l_2}^Y |\mathcal{W}^{XY}_{l_1 l_2 L}|^2 \, .
\end{equation}
Denoting an unnormalised estimator by an overbar, $\bar{\phi}^{XY}_{LM} =
\mathcal{R}_L^{XY} \hat{\phi}_{LM}^{XY}$, the disconnected
contribution to its power spectrum is $\langle \bar{\phi}_{LM}^{XY}
[\bar{\phi}_{L'M'}^{X'Y'}]^\ast\rangle_G = \bar{N}_L^{(0)}(XY,X'Y')
\delta_{LL'} \delta_{MM'}$ where
\begin{multline}
\bar{N}_L^{(0)}(XY,X'Y') = \frac{1}{4(2L+1)} \sum_{l_1 l_2}
[\mathcal{W}^{XY}_{l_1 l_2 L}]^\ast F_{l_1}^X F_{l_2}^Y \left(
\mathcal{W}^{X'Y'}_{l_1 l_2 L}F_{l_1}^{X'} F_{l_2}^{Y'}
C_{l_1,\text{tot}}^{XX'} C_{l_2,\text{tot}}^{YY'} \right. \\ \left.
+ \mathcal{W}^{Y'X'}_{l_1 l_2 L}F_{l_1}^{Y'} F_{l_2}^{X'}
C_{l_1,\text{tot}}^{XY'} C_{l_2,\text{tot}}^{YX'} \right) \, .
\end{multline}
The minimum-variance combination of (a subset of) the individual quadratic
estimators is approximately $\hat{\phi}^{MV}_{LM} = \sum_{XY}
\bar{\phi}_{LM}^{XY} / \sum_{XY} \mathcal{R}_L^{XY}$, and has
reconstruction noise power
\begin{equation}
N_L^{(0)}(MV) = \frac{1}{\left(\sum_{XY}
    \mathcal{R}_L^{XY}\right)^2}\sum_{XY} \sum_{X'Y'}
\bar{N}_L^{(0)}(XY,X'Y') \, .
\end{equation}
%


\acknowledgments

AC and RA acknowledge support from the UK Science and Technology
Facilities Council (grant number ST/N000927/1) as does AL (grant number
ST/L000652/1). AL and JC acknowledge support from the European Research Council under
the European Union's Seventh Framework Programme (FP/2007-2013) / ERC
Grant Agreement No. [616170].
J.G.N. acknowledges financial support from the Spanish MINECO for a `Ramon y Cajal' Fellowship (RYC-2013-13256) and the I+D 2015 project AYA2015-65887-P (MINECO/FEDER)."
CJM is supported by an FCT Research Professorship, contract reference
IF/00064/2012, funded by FCT/MCTES (Portugal) and POPH/FSE (EC).
Some of the results in this paper have
been derived using the {\tt HEALPix} package~\cite{Gorski:2004by}.

\bibliography{lensing}
\bibliographystyle{JHEP}

\end{document}

%% file: macros.tex
\def\WMAP{{WMAP}}

\newcommand{\mksym}[1]{\ifmmode {\rm #1}\else #1\fi}

\newcommand{\nnu}{N_{\rm eff}}

\newcommand{\mnusterile}{m_{\nu,\, \mathrm{sterile}}^{\mathrm{eff}}}
\newcommand{\meffsterile}{\mnusterile}

\providecommand{\Planck}{\textit{Planck}}
\providecommand{\planck}{\Planck}

\providecommand{\text}[1]{\rm{#1}}

\newcommand{\begm}{\begin{pmatrix}}
\newcommand{\enm}{\end{pmatrix}}

\newcommand\ba{\begin{eqnarray}}
\newcommand\ea{\end{eqnarray}}
\newcommand\bea{\begin{eqnarray}}
\newcommand\eea{\end{eqnarray}}

\newcommand\be{\begin{equation}}
\newcommand\ee{\end{equation}}

\def\pmb#1{\setbox0=\hbox{#1}%
    \kern-.025em\copy0\kern-\wd0
    \kern.05em\copy0\kern-\wd0
    \kern-.025em\raise.0433em\box0}

\def\p2Y{\;_2Y}
\def\m2Y{\;_{-2}Y}
\def\beglet{
  \addtocounter{equation}{1}%
  \setcounter{parentequation}{\value{equation}}%
  \setcounter{equation}{0}%
  \def\theequation{\arabic{parentequation}\alph{equation}}%
  \ignorespaces
}
\def\endlet{
  \setcounter{equation}{\value{parentequation}}%
  \def\theequation{\arabic{equation}}%
}
\providecommand{\beglet}{\begin{subequations}}
\providecommand{\endlet}{\end{subequations}}